\documentclass[aps,prl,reprint,superscriptaddress,floatfix]{revtex4-2}
\usepackage[dvipdfmx]{graphicx}
\usepackage{color}
\usepackage{ulem}
\begin{document}

\title{Broken Screw Rotational Symmetry in the Near-Surface Electronic Structure of $AB$-Stacked Crystals}

\author{Hiroaki~Tanaka}
\affiliation{Institute for Solid State Physics, The University of Tokyo, Kashiwa, Chiba 277-8581, Japan}

\author{Shota~Okazaki}
\affiliation{Materials and Structures Laboratory, Tokyo Institute of Technology, Yokohama, Kanagawa 226-8503, Japan}

\author{Masaru~Kobayashi}
\affiliation{Materials and Structures Laboratory, Tokyo Institute of Technology, Yokohama, Kanagawa 226-8503, Japan}

\author{Yuto~Fukushima}
\affiliation{Institute for Solid State Physics, The University of Tokyo, Kashiwa, Chiba 277-8581, Japan}

\author{Yosuke~Arai}
\affiliation{Institute for Solid State Physics, The University of Tokyo, Kashiwa, Chiba 277-8581, Japan}

\author{Takushi~Iimori}
\affiliation{Institute for Solid State Physics, The University of Tokyo, Kashiwa, Chiba 277-8581, Japan}

\author{Mikk~Lippmaa}
\affiliation{Institute for Solid State Physics, The University of Tokyo, Kashiwa, Chiba 277-8581, Japan}

\author{Kohei~Yamagami}
\affiliation{Japan Synchrotron Radiation Research Institute (JASRI), Sayo, Hyogo 679-5198, Japan}

\author{Yoshinori~Kotani}
\affiliation{Japan Synchrotron Radiation Research Institute (JASRI), Sayo, Hyogo 679-5198, Japan}

\author{Fumio~Komori}
\affiliation{Institute of Industrial Science, The University of Tokyo, Meguro-ku, Tokyo 153-8505, Japan}

\author{Kenta~Kuroda}
\email{kuroken224@hiroshima-u.ac.jp}
\affiliation{Graduate School of Advanced Science and Engineering, Hiroshima University, Higashi-hiroshima, Hiroshima 739-8526, Japan}
\affiliation{International Institute for Sustainability with Knotted Chiral Meta Matter (WPI-SKCM${}^{2}$), Higashi-hiroshima, Hiroshima 739-8526, Japan}

\author{Takao~Sasagawa}
\email{sasagawa@msl.titech.ac.jp}
\affiliation{Materials and Structures Laboratory, Tokyo Institute of Technology, Yokohama, Kanagawa 226-8503, Japan}

\author{Takeshi~Kondo}
\email{kondo1215@issp.u-tokyo.ac.jp}
\affiliation{Institute for Solid State Physics, The University of Tokyo, Kashiwa, Chiba 277-8581, Japan}
\affiliation{Trans-scale Quantum Science Institute, The University of Tokyo, Bunkyo-ku, Tokyo 113-0033, Japan}

\date{\today}

\begin{abstract}
We investigate the electronic structure of $2H$-$\mathrm{Nb}\mathrm{S}_2$ and $h$-BN by angle-resolved photoemission spectroscopy (ARPES) and photoemission intensity calculations.
Although in bulk form, these materials are expected to exhibit band degeneracy in the $k_z=\pi/c$ plane due to screw rotation and time-reversal symmetries, we observe gapped band dispersion near the surface.
We extract from first-principles calculations the near-surface electronic structure probed by ARPES and find that the calculated photoemission spectra from the near-surface region reproduce the gapped ARPES spectra.
Our results show that the near-surface electronic structure can be qualitatively different from the bulk one due to partially broken nonsymmorphic symmetries.
\end{abstract}

\maketitle

%The estimate is [(150 / aspect ratio) + 20 words] for single-column figures, and [300 / (0.5 * aspect ratio)] + 40 words for double-column figures.

Nonsymmorphic symmetries, such as screw rotations and glide reflections, can induce characteristic degeneracies in the band dispersions of crystals.
One such degeneracy is a nodal surface that is stabilized by a combination of a $2_1$ screw rotation with time-reversal symmetry.
If spin-orbit coupling (SOC) can be ignored, these two symmetries lead to degenerate band pairs at Brillouin zone boundary surfaces \cite{Chang2018}.
A gap can form at such nodal planes in the presence of SOC, but the band degeneracy is partially protected by the nonsymmorphic symmetries, forming nodal lines \cite{PhysRevB.93.085427, Schoop2016, doi:10.7566/JPSJ.88.044711}.
Recent studies have shown that such nodal line materials can exhibit extremely large magnetoresistance \cite{doi:10.1063/1.4953772, doi:10.1126/sciadv.1601742, doi:10.1073/pnas.1618004114} and other emergent properties, raising the prospect of device applications.

Furthermore, since the nonsymmorphic symmetries responsible for the formation of the nodal lines may not be maintained in the surface region, the electronic structure near the surface may differ from the bulk one.
For example, a nodal line material ZrSiS has been reported to exhibit surface-specific states \cite{PhysRevX.7.041073}.
Angle-resolved photoemission spectroscopy (ARPES) is a suitable probe for investigating the near-surface electronic structure because its probing depth is up to a few nanometers from the surface due to electron scattering processes \cite{https://doi.org/10.1002/sia.740010103}.
In addition, it has been unclear whether bulk band degeneracy due to nonsymmorphic symmetries can be retained in the near-surface region.
While multifold degeneracy has been observed by ARPES in nonsymmorphic CoSi \cite{PhysRevLett.122.076402, Rao2019}, no degeneracy was observed in a nodal line material $2H$-$\mathrm{Nb}\mathrm{S}_2$ \cite{PhysRevB.105.L121102}.
Clearly, more detailed experiments and calculations are needed to interpret the ARPES results of nonsymmorphic materials.

In this Letter, we present an extensive soft-x-ray (SX) ARPES study of the differences between the near-surface electronic structure and the bulk structure hosting nodal lines.
We examine the near-surface electronic structure of $AB$-stacked $2H$-$\mathrm{Nb}\mathrm{S}_2$ and $h$-BN (space group $P6_3/mmc$) crystals, both of which host bulk nodal lines on the $k_z=\pi/c$ plane.
Since $h$-BN is a wide-gap semiconductor, we used thin flakes of $h$-BN to avoid charge-up during ARPES measurements.
Although ARPES measurements of $h$-BN flakes are technically difficult, we chose $h$-BN because fewer valence electrons in $h$-BN compared to $2H$-$\mathrm{Nb}\mathrm{S}_2$ make the spectrum analysis easier and give more comprehensive evidence.
We found gapped band dispersions on the $k_z=\pi/c$ plane, indicating that the broken screw rotational symmetry at the surface opens an energy gap in the near-surface electronic structure.
Such band splitting cannot be explained by the $k_z$-broadening effect with the typical length scale $\Delta k_z=0.1\ \text{\AA}^{-1}$ \cite{RevModPhys.93.025006}, which has been frequently employed to discuss the surface sensitivity of ARPES.
Taking into account the surface sensitivity, our photoemission intensity calculations could reproduce such gapped spectra by adjusting the probing depth parameter in the simulation.
Our results show that incomplete nonsymmorphic symmetries can alter the near-surface electronic structure associated with the bulk nodal lines, not only inducing emergent surface states as previously reported \cite{PhysRevX.7.041073}.

Single crystals of $2H$-$\mathrm{Nb}\mathrm{S}_2$ were grown by the chemical vapor transport method, as explained in an earlier report \cite{PhysRevB.105.L121102}.
Single crystals of $h$-BN were grown by the flux method under atmospheric pressure; information on crystal growth and characterization is in the Supplemental Material Note 1 \cite{Supplemental}.
SX-ARPES measurements were performed at BL25SU of SPring-8 \cite{Muro:ok5049}, using 370 -- 760~eV SX light.
The measurement temperature was kept at around 50 K, and the energy resolution was $\approx$ 80 meV.
Since $2H$-$\mathrm{Nb}\mathrm{S}_2$ is a metal with sufficient conductivity, we cleaved as-grown samples \textit{in situ} in an ultrahigh vacuum better than $\sim 3\times10^{-8}$~Pa to obtain clean surfaces.
On the other hand, special care was taken in the preparation of $h$-BN to avoid charge-up.
We used thin exfoliated flakes of $h$-BN to reduce the resistance between the substrate and the top surface of the sample.
Note 1 in \cite{Supplemental} explains the detailed procedure of the flake fabrication and the dry pickup method of sample mounting in the ARPES chamber \cite{doi:10.1021/acs.nanolett.8b04534, Masubuchi2022}.
The $h$-BN flakes were several tens of micrometers wide and a few hundred nanometers thick [Fig.\ S2 in \cite{Supplemental}].
Since these samples are much smaller than typical as-grown crystals, we used micrometer-focused SX \cite{Muro:ok5049}.
In the photoemission intensity calculations, we used OpenMX \cite{PhysRevB.67.155108} and a recently developed photoemission angular distribution simulator SPADExp \cite{TANAKA2023147297}.
Note 2 in \cite{Supplemental} summarizes the calculation conditions.

% 86 mm x 122 mm -> 233 words
% caption: 120 words
\begin{figure}
\includegraphics{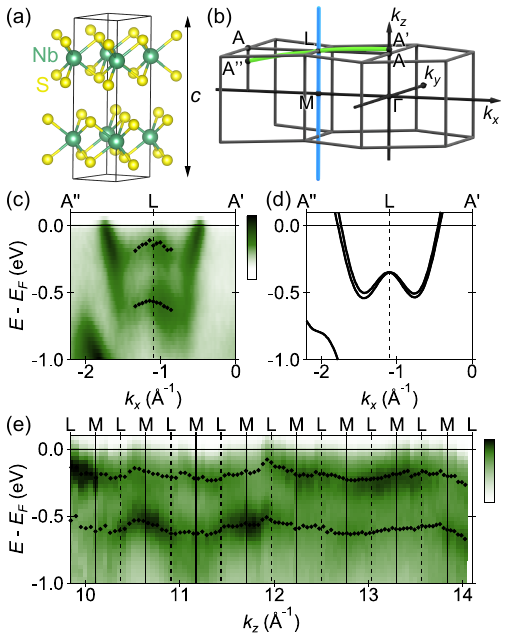} 
\caption{\label{Fig: NbS2_kz} Band dispersion of $2H$-$\mathrm{Nb}\mathrm{S}_2$. (a) Crystal structure. The unit cell height is $c=11.89$~\AA. (b) The Brillouin zone of a hexagonal crystal with high-symmetry point labels. The blue and green curves represent momentum paths along which the band dispersions are presented. The green curve corresponds to the momentum path in single-photon-energy ARPES measurements. (c) Band dispersion along the green curve taken with 525~eV SX light. The $L$ point is on the $k_z=22.5\times2\pi/c$ plane. (d) Bulk band dispersion calculated along the same path as (c). (e) $k_z$ dispersion along the blue $ML$ path. In (c) and (e), the black dots represent the band positions determined by fitting the energy distribution curves [Figs.\ S4(b) and S4(c) in \cite{Supplemental}, respectively].}
\end{figure}

We start our discussion with SX-ARPES results of $2H$-$\mathrm{Nb}\mathrm{S}_2$.
While a previous study has already reported the gapped band dispersion at the $L$ points \cite{PhysRevB.105.L121102}, we performed SX-ARPES measurements over a wider $k_z$ range to check the universality of this phenomenon.
The $AL$ path is suitable for our discussion because the crystal symmetry prohibits the gap opening due to SOC [Fig.\ S4(a) in \cite{Supplemental}].
Figure \ref{Fig: NbS2_kz}(c) clearly shows that one paired band at the valence band top has an energy gap of about 0.5~eV at the $L$ point.
On the other hand, the bulk band dispersion at the $L$ point exhibits symmetry-related degeneracy [Fig.\ \ref{Fig: NbS2_kz}(d)]; a small energy gap at the momentum points other than the $L$ point is due to the curved momentum path in single-photon-energy ARPES measurements  [Fig.\ \ref{Fig: NbS2_kz}(b)].
The $k_z$ dispersion along the $ML$ line, taken by changing the photon energy from 370 to 750~eV, also shows gapped dispersion at all observed $L$ points.
These results clearly show that the bulk band degeneracy caused by the screw rotational symmetry is broken in the near-surface electronic structure of $2H$-$\mathrm{Nb}\mathrm{S}_2$.

% 86 mm x 147 mm -> 276 words
% caption: 116 words
\begin{figure}
\includegraphics{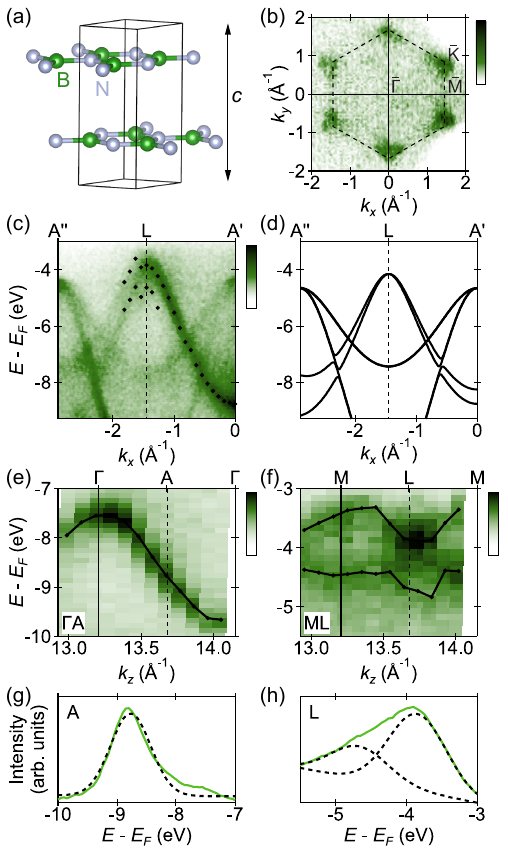} 
\caption{\label{Fig: hBN_kz} Band dispersion of $h$-BN. (a) Crystal structure. The unit cell height is $c=6.66$~\AA. (b) Constant-energy surface at $E_F-3$~eV taken with 750~eV SX light. The dashed hexagon represents the Brillouin zone. (c) Band dispersion crossing the $L$ point taken with 720 eV SX light. The $L$ point is on the $k_z=14.5\times2\pi/c$ plane. (d) Bulk band dispersion along the same path as (c). (e), (f) $k_z$ dispersions along the $\Gamma A$ and $ML$ paths, respectively. 
%The black dashed curve in (e) represents the absent band estimated by inverting the observed band along the horizontal axis. 
In (c), (e), and (f), the black dots and solid curves represent band positions determined by fitting the energy distribution curves [Figs.\ S7, S6(a), and S6(g) in \cite{Supplemental}, respectively]. (g), (h) Energy distribution curves at the $A$ and $L$ points, respectively. The black dashed curves represent the peaks included in the fitting function.}
%\caption{\label{Fig: hBN_kz}}
\end{figure}

Next, we investigated the electronic structure of $h$-BN [Fig.\ \ref{Fig: hBN_kz}(a)].
As was noted earlier, ARPES spectra of $h$-BN were easier to discuss due to fewer valence electrons in this material.
The use of the micrometer-focused SX beam ensured that the photoemission spectra were derived only from the $h$-BN flakes.
We observed the constant-energy surface [Fig.\ \ref{Fig: hBN_kz}(b)], which coincided well with a previous ARPES study of $h$-BN with sub-micrometer-focused vacuum ultraviolet (VUV) light \cite{PhysRevB.95.085410}, and signal from the substrate was negligible.
The core-level spectra clearly exhibited the $1s$ peaks of boron and nitrogen only when the soft-x-ray beam was on the $h$-BN flake [Fig.\ S5 in \cite{Supplemental}].
The band dispersion around the $L$ point has an energy gap of about 0.8~eV [Figs.\ \ref{Fig: hBN_kz}(c) and \ref{Fig: hBN_kz}(h)], contrary to the gapless bulk band dispersion [Fig.\ \ref{Fig: hBN_kz}(d)].
Such inconsistent behavior between experiment and bulk band calculations is similar to our SX-ARPES results of $2H$-$\mathrm{Nb}\mathrm{S}_2$ [Fig.\ \ref{Fig: NbS2_kz}].

Because of negligible SOC and fewer electrons, we could examine the degeneracy of the $h$-BN $\pi$ band along the entire $AL$ path.
We found that the ARPES behavior at the $A$ point differed from that at the $L$ point.

First, we observed only a single peak in the energy distribution curve extracted along the $\Gamma A$ path [Fig.\ S6(a) in \cite{Supplemental}].
Since the peak profiles are symmetric and fit well with a single Gaussian, they can be assigned to a single band.
As a result, the $k_z$ dispersion of the $\pi$ band along the $\Gamma A$ direction shows only a single oscillating dispersion profile, while the other one forming a pair is absent [Fig.\ \ref{Fig: hBN_kz}(e)].
Such a phenomenon has been frequently observed in the $k_z$ dispersion of $AB$-stacked materials \cite{PhysRevLett.119.026403, PhysRevB.97.045430, PhysRevB.105.L121102}.
If the $A$ and $B$ layers are similar, the electronic structure of the crystal can be approximated by a system with a height of $c/2$ containing only one layer.
The approximated electronic structure gives a single band with a periodicity of $2\pi/(c/2)=4\pi/c$.
We reproduced the $4\pi/c$ dispersion in the photoemission intensity calculations of the $h$-BN bulk [Fig.\ S10 in \cite{Supplemental}].

Second, more importantly, the $k_z$ dispersion along the $\Gamma A$ path has no energy gap at the $A$ point [Figs.\ \ref{Fig: hBN_kz}(e) and \ref{Fig: hBN_kz}(g)].
This behavior differs from the dispersion along the $ML$ path [Figs.\ \ref{Fig: hBN_kz}(f) and \ref{Fig: hBN_kz}(h)] and that of $2H$-$\mathrm{Nb}\mathrm{S}_2$ [Fig.\ \ref{Fig: NbS2_kz}(e)] but is consistent with ARPES study of graphite \cite{PhysRevB.97.045430}, which has a similar crystal structure to $h$-BN.
With increasing $k_x$ value from the $\Gamma A$ to $ML$ paths, we observed a crossover from the gapless ``bulk'' spectra to the gapped ``surface'' spectra [Note 5 in \cite{Supplemental}].
This crossover direction is physically reasonable; a nonzero $k_x$ component means that the photoelectron momentum vector is canted away from the surface normal, making the effective distance to the surface longer and the measurement more surface sensitive.
To summarize, ARPES spectra and bulk band dispersion can be substantially different at least partially, even if we perform relatively bulk-sensitive soft-x-ray ARPES \cite{https://doi.org/10.1002/sia.740010103}.
The deviation has been discussed as the $k_z$-broadening effect in the reciprocal space.
While it can explain the peak position shift \cite{STROCOV200365} and may cause the band splitting, our $k_z$-broadening simulation revealed that the typical broadening length scale $\Delta k_z=0.1\ \text{\AA}^{-1}$ \cite{RevModPhys.93.025006} was insufficient to cause the band splitting [Note 6 in \cite{Supplemental}].

\begin{figure}[th]
\includegraphics{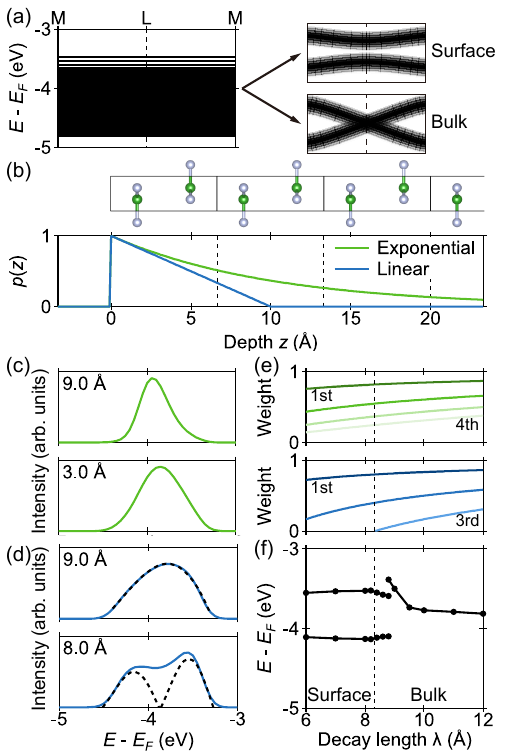} 
\caption{\label{Fig: hBN_PAD_EDCs} Photoemission intensity calculations of $h$-BN. (a) Calculated slab band dispersion along the $ML$ path and schematic surface and bulk spectra obtained by successive photoemission intensity calculations. (b) Exponential and linear decay functions, where the depth parameter $\lambda$ is set to 10~\AA. The dashed grid corresponds to the height of the unit cell, as drawn in the top panel. (c), (d) Calculated spectra at the $L$ point using the exponential and linear decay functions with different $\lambda$ values, respectively. In (d), the black dashed curves represent the peaks included in the fitting function. (e) $\lambda$ dependence of the decay strength at each atomic layer position for both exponential and linear functions. (f) $\lambda$ dependence of peak positions extracted from calculated spectra using the linear decay function [Fig.\ S11 in \cite{Supplemental}]. In (e) and (f), the dashed vertical lines represent the third layer position.}
\end{figure}

To elucidate the mechanism that leads to the gapped surface spectra, we performed photoemission intensity calculations taking into account the surface sensitivity of ARPES in real space.
The first-principles calculations of $h$-BN were done using a 40-bilayer slab to obtain the near-surface electronic structure.
However, the band dispersion includes the whole electronic structure of the slab and is very broad [Fig.\ \ref{Fig: hBN_PAD_EDCs}(a)].
We, therefore, performed photoemission intensity calculations to obtain the matrix element between the initial and final states and extract the near-surface electronic structure \cite{MOSER201729}.
The initial state is a ground state wave function obtained by first-principles calculations, and the final state is a plane wave with decay due to scattering.
Combining the large slab system and final states with decay enables the photoemission simulation to extract the near-surface electronic structure observed by ARPES \cite{TANAKA2023147297}.
If the decayed final states cover a sufficient number of unit cells along the $z$ direction, we can expect a bulk spectrum shaped by the momentum selection rule.
Otherwise, the calculated spectra will deviate from the bulk spectrum, resulting in a gapped surface spectrum [Fig.\ \ref{Fig: hBN_PAD_EDCs}(a)], as mainly observed in our ARPES measurements.
From this consideration, we can claim that the band splitting due to the broken nonsymmorphic symmetry is smaller than the band width along the $k_z$ direction.

% 86 mm x 134 mm -> 254 words
% caption: 179 words

We compared two functions for the square of decay related to the photoelectron emission probability $p(z)$; an exponential function $\exp(-z/\lambda)$ and a linear function $\max(1-z/\lambda, 0)$ [Fig.\ \ref{Fig: hBN_PAD_EDCs}(b)].
Exponential depth decay has been commonly used in discussing the surface sensitivity of ARPES \cite{MOSER201729} based on the inelastic scattering measurements of photoelectrons from polycrystalline films \cite{PhysRevB.1.522, M_Klasson_1972}.
However, it did not reproduce the gapped surface spectra even for unreasonably small mean free path parameter $\lambda=3$~{\AA} [Fig.\ \ref{Fig: hBN_PAD_EDCs}(c)].
The gapless spectrum at $\lambda=3$~{\AA} is consistent with our $k_z$-broadening simulation ($\Delta k_z= 0.6 \pi/c$) exhibiting band position shift but preserved band degeneracy [Fig.\ S9(c) in \cite{Supplemental}].
The disagreement between experiments and simulations may be because the exponential decay is too slow to reproduce the ARPES spectra, where both the inelastic and elastic scatterings matter.
Therefore, instead of the exponential decay, we employed the linear decay function, which is steeper than exponential and has the cutoff property.
The steeper decay function can be interpreted as a different peak shape than what has been used before in the context of the $k_z$-broadening effect \cite{STROCOV200365}.
The linear decay function is appropriate for comprehensive discussion because it is controlled only by one continuous parameter $\lambda$.
Using the linear decay function, we reproduced the gapped surface spectra that could be fitted well with two Gaussians when $\lambda< 8.8$~{\AA} [Fig.\ \ref{Fig: hBN_PAD_EDCs}(d)].
The crossover from a single to a double peak spectrum happens abruptly [Fig.\ S11 in \cite{Supplemental}], and the position where the change occurs seems to be related to the third layer from the surface.
Figures \ref{Fig: hBN_PAD_EDCs}(e) and \ref{Fig: hBN_PAD_EDCs}(f) show that when we increase the depth parameter $\lambda$, the surface spectra abruptly change to the bulk shape as the decay function starts to cover the third layer, marked with the dashed vertical line in the figures.
The exponential decay function, therefore, always produced a gapless bulk spectrum due to the long decay tail that covers deeper layers, even if $\lambda$ is quite small [Fig.\ \ref{Fig: hBN_PAD_EDCs}(e) top panel].
While the linear decay function successfully reproduced gapped spectra, the decay shape is empirical.
We hope further research will discover the decay function shape due to inelastic and elastic scattering by calculations and/or experiments.

% 86 mm x 88 mm -> 173 words
% caption: 62 words
\begin{figure}
\includegraphics{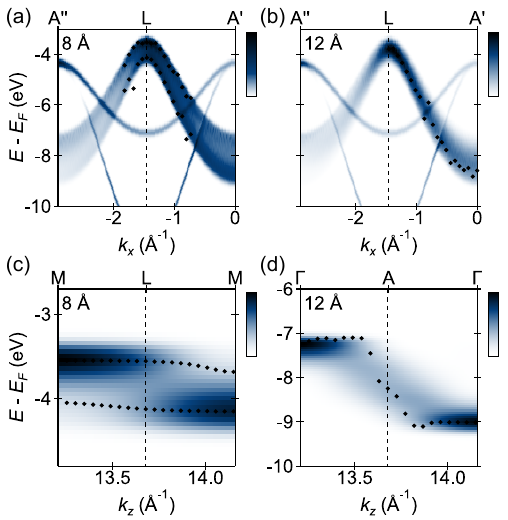} 
\caption{\label{Fig: hBN_PAD_maps} Photoemission angular distribution calculations of $h$-BN. (a), (b) Calculated band dispersions along the path crossing the $L$ point. (c), (d) Calculated band dispersions along the $ML$ and $\Gamma A$ paths, respectively. The left top labels represent the linear decay parameter $\lambda$. The black dots in the panels represent peak positions determined by fitting the energy distribution curves [Figs.\ S12 and S13 in \cite{Supplemental}].}
\end{figure}

Finally, we performed photoemission angular distribution calculations for $h$-BN to validate the SX-ARPES results [Fig.\ \ref{Fig: hBN_kz}].
The calculated photoemission spectra around the $L$ point, using a linear decay function with $\lambda=8$~{\AA} shows an energy gap [Fig.\ \ref{Fig: hBN_PAD_maps}(a)] that is very similar to the experimental observation [Fig.\ \ref{Fig: hBN_kz}(c)].
On the other hand, using just a slightly larger decay parameter of $\lambda=12$~{\AA} produces a gapless dispersion curve [Fig.\ \ref{Fig: hBN_PAD_maps}(b)], agreeing well with the experimental data around the $A$ point.
The $k_z$ dispersions along the $ML$ and $\Gamma A$ paths reproduce the gapped and gapless behaviors [Figs.\ \ref{Fig: hBN_PAD_maps}(c) and (d)] when an appropriate decay parameter is chosen.
These results indicate that the out-of-plane scattering length slightly depends on the in-plane momentum of a photoelectron, even for the same $\pi$ band.
This assumption is reasonable because the negative correlation between the in-plane momentum and the scattering length is physically natural, as explained earlier in the paragraph on $h$-BN experiments, and the scattering length difference needed to explain the different experimental spectra is not large.
In addition, the nonuniform electronic structure in single crystals may enhance the angle dependence of the scattering phenomena.

Additionally, the decay conditions used in the $h$-BN calculations succeeded in reproducing the gapped surface spectra of $2H$-$\mathrm{Nb}\mathrm{S}_2$ [Fig.\ S14 in \cite{Supplemental}].
Since the third layer of $2H$-$\mathrm{Nb}\mathrm{S}_2$ is deeper than $h$-BN, the photoemission spectra are always gapped, independent of the selection of the decay length parameter.
We can exclude the possibility of a SOC-derived energy gap because the gap behavior is independent of the inclusion of SOC.
Our results indicate that photoemission intensity simulations using a linear decay function can be universally applicable to other materials because the photoelectron inelastic mean free path exhibits a universal energy dependence \cite{https://doi.org/10.1002/sia.740010103}.
In addition, our $\lambda$ dependence analysis of the simulated spectra gives insight into previous observations of multifold degeneracy in nonsymmorphic CoSi \cite{PhysRevLett.122.076402, Rao2019} and Dirac points in $\mathrm{Na}_3\mathrm{Bi}$ \cite{doi:10.1126/science.1245085}.
Since CoSi has a three-dimensional structure, which can be interpreted as a stacked material composed of very thin layers, more layers are included within the ARPES probing depth than in the case of quasi-two-dimensional $2H$-$\mathrm{Nb}\mathrm{S}_2$ and $h$-BN.
Considering our discussion that the number of layers included in the probing depth is essential to determine the appearance of bulk degeneracy in ARPES spectra, three-dimensional nonsymmorphic materials seem to be more likely to exhibit bulk band degeneracy in ARPES experiments.
On the other hand, since the Dirac points in $\mathrm{Na}_3\mathrm{Bi}$ are protected by in-plane rotational symmetry, their degeneracy will be conserved in the near-surface electronic structure, consistent with the clear observation of the Dirac points by ARPES \cite{doi:10.1126/science.1245085}.
Nevertheless, another possibility is that their observed degeneracy can be decomposed into several peaks if the energy distribution curve is analyzed in more detail, as we did in the present study.
In addition, since the observed degeneracies discussed previously are located near the Fermi level, other peak structures may exist above the Fermi level and therefore be absent in experimental spectra.

In conclusion, we investigated the symmetry-protected band degeneracy in $2H$-$\mathrm{Nb}\mathrm{S}_2$ and $h$-BN by ARPES and photoemission intensity calculations.
While the bulk band degeneracy in these materials is due to the coexistence of $2_1$ screw rotation and time-reversal symmetries, the near-surface electronic structure that ARPES can probe may differ from the bulk because the screw rotation symmetry is broken at the surface.
Indeed, we observed gapped band dispersion on the $k_z=\pi/c$ plane in most cases other than around the $A$ point of $h$-BN.
Our photoemission intensity calculations to extract the near-surface electronic structure revealed that the assumption of exponential photoemission probability decay, which has been frequently used to discuss the surface sensitivity of ARPES, cannot reproduce the experimental results.
Instead, a steeper linear decay reproduced the gapped behavior with an abrupt crossover to gapless bulk spectra when the decay length parameter was increased, giving a reasonable explanation for the simultaneous observation of surface and bulk spectra at the $L$ and $A$ points of $h$-BN.
This work demonstrates the necessity to consider the surface electronic structure to discuss photoemission spectra, even in relatively bulk-sensitive SX-ARPES.
Furthermore, we show that the electronic structure in the near-surface region can deviate significantly from the symmetry-protected bulk one, providing a possibility of a more exotic electronic structure in the surface or interface of nodal line materials.

\begin{acknowledgments}
We thank Abdelkarim Ouerghi for providing their experimental methods of VUV-ARPES measurements of $h$-BN \cite{PhysRevB.95.085410} and Haruki Watanabe for theoretical advice.
This work was also supported by Grant-in-Aid for JSPS Fellows (Grant No.\ JP21J20657), Grant-in-Aid for Scientific Research on Innovative Areas (Grants No.\ JP21H05236 and No.\ JP22H04483), Grant-in-Aid for Challenging Research (Pioneering) (Grants No.\ JP21K18181 No.\ JP23K17351), Grant-in-Aid for Scientific Research (B) (Grant No.\ JP22H01943) Grant-in-Aid for Scientific Research (A) (Grants No.\ JP21H04652 and No.\ JP21H04439), Japan Science and Technology Agency (JST) (Grant No.\ JPMJMI21G2), Quantum Leap Flagship Program from Ministry of Education, Culture, Sports, Science and Technology (MEXT Q-LEAP) (Grant No.\ JPMXS0118068681), the Asahi Glass Foundation, the Murata Science Foundation, and Photon and Quantum Basic Research Coordinated Development Program from MEXT.
Optical microscope measurements were performed using the facilities of the Materials Design and Characterization Laboratory in the Institute for Solid State Physics, the University of Tokyo.
The synchrotron radiation experiments were performed with the approval of Japan Synchrotron Radiation Research Institute (JASRI) (Proposals No.\ 2020A1181, No.\ 2021A1259, No.\ 2021B1797, and No.\ 2022A1687).
\end{acknowledgments}
\bibliography{hBN_NbS2_ARPES_references}

\begin{thebibliography}{27}%
\makeatletter
\providecommand \@ifxundefined [1]{%
 \@ifx{#1\undefined}
}%
\providecommand \@ifnum [1]{%
 \ifnum #1\expandafter \@firstoftwo
 \else \expandafter \@secondoftwo
 \fi
}%
\providecommand \@ifx [1]{%
 \ifx #1\expandafter \@firstoftwo
 \else \expandafter \@secondoftwo
 \fi
}%
\providecommand \natexlab [1]{#1}%
\providecommand \enquote  [1]{``#1''}%
\providecommand \bibnamefont  [1]{#1}%
\providecommand \bibfnamefont [1]{#1}%
\providecommand \citenamefont [1]{#1}%
\providecommand \href@noop [0]{\@secondoftwo}%
\providecommand \href [0]{\begingroup \@sanitize@url \@href}%
\providecommand \@href[1]{\@@startlink{#1}\@@href}%
\providecommand \@@href[1]{\endgroup#1\@@endlink}%
\providecommand \@sanitize@url [0]{\catcode `\\12\catcode `\$12\catcode
  `\&12\catcode `\#12\catcode `\^12\catcode `\_12\catcode `\%12\relax}%
\providecommand \@@startlink[1]{}%
\providecommand \@@endlink[0]{}%
\providecommand \url  [0]{\begingroup\@sanitize@url \@url }%
\providecommand \@url [1]{\endgroup\@href {#1}{\urlprefix }}%
\providecommand \urlprefix  [0]{URL }%
\providecommand \Eprint [0]{\href }%
\providecommand \doibase [0]{https://doi.org/}%
\providecommand \selectlanguage [0]{\@gobble}%
\providecommand \bibinfo  [0]{\@secondoftwo}%
\providecommand \bibfield  [0]{\@secondoftwo}%
\providecommand \translation [1]{[#1]}%
\providecommand \BibitemOpen [0]{}%
\providecommand \bibitemStop [0]{}%
\providecommand \bibitemNoStop [0]{.\EOS\space}%
\providecommand \EOS [0]{\spacefactor3000\relax}%
\providecommand \BibitemShut  [1]{\csname bibitem#1\endcsname}%
\let\auto@bib@innerbib\@empty
%</preamble>
\bibitem [{\citenamefont {Chang}\ \emph {et~al.}(2018)\citenamefont {Chang},
  \citenamefont {Wieder}, \citenamefont {Schindler}, \citenamefont {Sanchez},
  \citenamefont {Belopolski}, \citenamefont {Huang}, \citenamefont {Singh},
  \citenamefont {Wu}, \citenamefont {Chang}, \citenamefont {Neupert},
  \citenamefont {Xu}, \citenamefont {Lin},\ and\ \citenamefont
  {Hasan}}]{Chang2018}%
  \BibitemOpen
  \bibfield  {author} {\bibinfo {author} {\bibfnamefont {G.}~\bibnamefont
  {Chang}}, \bibinfo {author} {\bibfnamefont {B.~J.}\ \bibnamefont {Wieder}},
  \bibinfo {author} {\bibfnamefont {F.}~\bibnamefont {Schindler}}, \bibinfo
  {author} {\bibfnamefont {D.~S.}\ \bibnamefont {Sanchez}}, \bibinfo {author}
  {\bibfnamefont {I.}~\bibnamefont {Belopolski}}, \bibinfo {author}
  {\bibfnamefont {S.-M.}\ \bibnamefont {Huang}}, \bibinfo {author}
  {\bibfnamefont {B.}~\bibnamefont {Singh}}, \bibinfo {author} {\bibfnamefont
  {D.}~\bibnamefont {Wu}}, \bibinfo {author} {\bibfnamefont {T.-R.}\
  \bibnamefont {Chang}}, \bibinfo {author} {\bibfnamefont {T.}~\bibnamefont
  {Neupert}}, \bibinfo {author} {\bibfnamefont {S.-Y.}\ \bibnamefont {Xu}},
  \bibinfo {author} {\bibfnamefont {H.}~\bibnamefont {Lin}},\ and\ \bibinfo
  {author} {\bibfnamefont {M.~Z.}\ \bibnamefont {Hasan}},\ }\href
  {https://doi.org/10.1038/s41563-018-0169-3} {\bibfield  {journal} {\bibinfo
  {journal} {Nature Materials}\ }\textbf {\bibinfo {volume} {17}},\ \bibinfo
  {pages} {978} (\bibinfo {year} {2018})}\BibitemShut {NoStop}%
\bibitem [{\citenamefont {Liang}\ \emph {et~al.}(2016)\citenamefont {Liang},
  \citenamefont {Zhou}, \citenamefont {Yu}, \citenamefont {Wang},\ and\
  \citenamefont {Weng}}]{PhysRevB.93.085427}%
  \BibitemOpen
  \bibfield  {author} {\bibinfo {author} {\bibfnamefont {Q.-F.}\ \bibnamefont
  {Liang}}, \bibinfo {author} {\bibfnamefont {J.}~\bibnamefont {Zhou}},
  \bibinfo {author} {\bibfnamefont {R.}~\bibnamefont {Yu}}, \bibinfo {author}
  {\bibfnamefont {Z.}~\bibnamefont {Wang}},\ and\ \bibinfo {author}
  {\bibfnamefont {H.}~\bibnamefont {Weng}},\ }\href
  {https://doi.org/10.1103/PhysRevB.93.085427} {\bibfield  {journal} {\bibinfo
  {journal} {Phys. Rev. B}\ }\textbf {\bibinfo {volume} {93}},\ \bibinfo
  {pages} {085427} (\bibinfo {year} {2016})}\BibitemShut {NoStop}%
\bibitem [{\citenamefont {Schoop}\ \emph {et~al.}(2016)\citenamefont {Schoop},
  \citenamefont {Ali}, \citenamefont {Stra{\ss}er}, \citenamefont {Topp},
  \citenamefont {Varykhalov}, \citenamefont {Marchenko}, \citenamefont
  {Duppel}, \citenamefont {Parkin}, \citenamefont {Lotsch},\ and\ \citenamefont
  {Ast}}]{Schoop2016}%
  \BibitemOpen
  \bibfield  {author} {\bibinfo {author} {\bibfnamefont {L.~M.}\ \bibnamefont
  {Schoop}}, \bibinfo {author} {\bibfnamefont {M.~N.}\ \bibnamefont {Ali}},
  \bibinfo {author} {\bibfnamefont {C.}~\bibnamefont {Stra{\ss}er}}, \bibinfo
  {author} {\bibfnamefont {A.}~\bibnamefont {Topp}}, \bibinfo {author}
  {\bibfnamefont {A.}~\bibnamefont {Varykhalov}}, \bibinfo {author}
  {\bibfnamefont {D.}~\bibnamefont {Marchenko}}, \bibinfo {author}
  {\bibfnamefont {V.}~\bibnamefont {Duppel}}, \bibinfo {author} {\bibfnamefont
  {S.~S.~P.}\ \bibnamefont {Parkin}}, \bibinfo {author} {\bibfnamefont {B.~V.}\
  \bibnamefont {Lotsch}},\ and\ \bibinfo {author} {\bibfnamefont {C.~R.}\
  \bibnamefont {Ast}},\ }\href {https://doi.org/10.1038/ncomms11696} {\bibfield
   {journal} {\bibinfo  {journal} {Nature Communications}\ }\textbf {\bibinfo
  {volume} {7}},\ \bibinfo {pages} {11696} (\bibinfo {year}
  {2016})}\BibitemShut {NoStop}%
\bibitem [{\citenamefont {Funada}\ \emph {et~al.}(2019)\citenamefont {Funada},
  \citenamefont {Yamakage}, \citenamefont {Yamashina},\ and\ \citenamefont
  {Kageyama}}]{doi:10.7566/JPSJ.88.044711}%
  \BibitemOpen
  \bibfield  {author} {\bibinfo {author} {\bibfnamefont {K.}~\bibnamefont
  {Funada}}, \bibinfo {author} {\bibfnamefont {A.}~\bibnamefont {Yamakage}},
  \bibinfo {author} {\bibfnamefont {N.}~\bibnamefont {Yamashina}},\ and\
  \bibinfo {author} {\bibfnamefont {H.}~\bibnamefont {Kageyama}},\ }\href
  {https://doi.org/10.7566/JPSJ.88.044711} {\bibfield  {journal} {\bibinfo
  {journal} {Journal of the Physical Society of Japan}\ }\textbf {\bibinfo
  {volume} {88}},\ \bibinfo {pages} {044711} (\bibinfo {year}
  {2019})}\BibitemShut {NoStop}%
\bibitem [{\citenamefont {Lv}\ \emph {et~al.}(2016)\citenamefont {Lv},
  \citenamefont {Zhang}, \citenamefont {Li}, \citenamefont {Yao}, \citenamefont
  {Chen}, \citenamefont {Zhou}, \citenamefont {Zhang}, \citenamefont {Lu},\
  and\ \citenamefont {Chen}}]{doi:10.1063/1.4953772}%
  \BibitemOpen
  \bibfield  {author} {\bibinfo {author} {\bibfnamefont {Y.-Y.}\ \bibnamefont
  {Lv}}, \bibinfo {author} {\bibfnamefont {B.-B.}\ \bibnamefont {Zhang}},
  \bibinfo {author} {\bibfnamefont {X.}~\bibnamefont {Li}}, \bibinfo {author}
  {\bibfnamefont {S.-H.}\ \bibnamefont {Yao}}, \bibinfo {author} {\bibfnamefont
  {Y.~B.}\ \bibnamefont {Chen}}, \bibinfo {author} {\bibfnamefont
  {J.}~\bibnamefont {Zhou}}, \bibinfo {author} {\bibfnamefont {S.-T.}\
  \bibnamefont {Zhang}}, \bibinfo {author} {\bibfnamefont {M.-H.}\ \bibnamefont
  {Lu}},\ and\ \bibinfo {author} {\bibfnamefont {Y.-F.}\ \bibnamefont {Chen}},\
  }\href {https://doi.org/10.1063/1.4953772} {\bibfield  {journal} {\bibinfo
  {journal} {Applied Physics Letters}\ }\textbf {\bibinfo {volume} {108}},\
  \bibinfo {pages} {244101} (\bibinfo {year} {2016})}\BibitemShut {NoStop}%
\bibitem [{\citenamefont {Ali}\ \emph {et~al.}(2016)\citenamefont {Ali},
  \citenamefont {Schoop}, \citenamefont {Garg}, \citenamefont {Lippmann},
  \citenamefont {Lara}, \citenamefont {Lotsch},\ and\ \citenamefont
  {Parkin}}]{doi:10.1126/sciadv.1601742}%
  \BibitemOpen
  \bibfield  {author} {\bibinfo {author} {\bibfnamefont {M.~N.}\ \bibnamefont
  {Ali}}, \bibinfo {author} {\bibfnamefont {L.~M.}\ \bibnamefont {Schoop}},
  \bibinfo {author} {\bibfnamefont {C.}~\bibnamefont {Garg}}, \bibinfo {author}
  {\bibfnamefont {J.~M.}\ \bibnamefont {Lippmann}}, \bibinfo {author}
  {\bibfnamefont {E.}~\bibnamefont {Lara}}, \bibinfo {author} {\bibfnamefont
  {B.}~\bibnamefont {Lotsch}},\ and\ \bibinfo {author} {\bibfnamefont
  {S.~S.~P.}\ \bibnamefont {Parkin}},\ }\href
  {https://doi.org/10.1126/sciadv.1601742} {\bibfield  {journal} {\bibinfo
  {journal} {Science Advances}\ }\textbf {\bibinfo {volume} {2}},\ \bibinfo
  {pages} {e1601742} (\bibinfo {year} {2016})}\BibitemShut {NoStop}%
\bibitem [{\citenamefont {Singha}\ \emph {et~al.}(2017)\citenamefont {Singha},
  \citenamefont {Pariari}, \citenamefont {Satpati},\ and\ \citenamefont
  {Mandal}}]{doi:10.1073/pnas.1618004114}%
  \BibitemOpen
  \bibfield  {author} {\bibinfo {author} {\bibfnamefont {R.}~\bibnamefont
  {Singha}}, \bibinfo {author} {\bibfnamefont {A.~K.}\ \bibnamefont {Pariari}},
  \bibinfo {author} {\bibfnamefont {B.}~\bibnamefont {Satpati}},\ and\ \bibinfo
  {author} {\bibfnamefont {P.}~\bibnamefont {Mandal}},\ }\href
  {https://doi.org/10.1073/pnas.1618004114} {\bibfield  {journal} {\bibinfo
  {journal} {Proceedings of the National Academy of Sciences}\ }\textbf
  {\bibinfo {volume} {114}},\ \bibinfo {pages} {2468} (\bibinfo {year}
  {2017})}\BibitemShut {NoStop}%
\bibitem [{\citenamefont {Topp}\ \emph {et~al.}(2017)\citenamefont {Topp},
  \citenamefont {Queiroz}, \citenamefont {Gr\"uneis}, \citenamefont
  {M\"uchler}, \citenamefont {Rost}, \citenamefont {Varykhalov}, \citenamefont
  {Marchenko}, \citenamefont {Krivenkov}, \citenamefont {Rodolakis},
  \citenamefont {McChesney}, \citenamefont {Lotsch}, \citenamefont {Schoop},\
  and\ \citenamefont {Ast}}]{PhysRevX.7.041073}%
  \BibitemOpen
  \bibfield  {author} {\bibinfo {author} {\bibfnamefont {A.}~\bibnamefont
  {Topp}}, \bibinfo {author} {\bibfnamefont {R.}~\bibnamefont {Queiroz}},
  \bibinfo {author} {\bibfnamefont {A.}~\bibnamefont {Gr\"uneis}}, \bibinfo
  {author} {\bibfnamefont {L.}~\bibnamefont {M\"uchler}}, \bibinfo {author}
  {\bibfnamefont {A.~W.}\ \bibnamefont {Rost}}, \bibinfo {author}
  {\bibfnamefont {A.}~\bibnamefont {Varykhalov}}, \bibinfo {author}
  {\bibfnamefont {D.}~\bibnamefont {Marchenko}}, \bibinfo {author}
  {\bibfnamefont {M.}~\bibnamefont {Krivenkov}}, \bibinfo {author}
  {\bibfnamefont {F.}~\bibnamefont {Rodolakis}}, \bibinfo {author}
  {\bibfnamefont {J.~L.}\ \bibnamefont {McChesney}}, \bibinfo {author}
  {\bibfnamefont {B.~V.}\ \bibnamefont {Lotsch}}, \bibinfo {author}
  {\bibfnamefont {L.~M.}\ \bibnamefont {Schoop}},\ and\ \bibinfo {author}
  {\bibfnamefont {C.~R.}\ \bibnamefont {Ast}},\ }\href
  {https://doi.org/10.1103/PhysRevX.7.041073} {\bibfield  {journal} {\bibinfo
  {journal} {Phys. Rev. X}\ }\textbf {\bibinfo {volume} {7}},\ \bibinfo {pages}
  {041073} (\bibinfo {year} {2017})}\BibitemShut {NoStop}%
\bibitem [{\citenamefont {Seah}\ and\ \citenamefont
  {Dench}(1979)}]{https://doi.org/10.1002/sia.740010103}%
  \BibitemOpen
  \bibfield  {author} {\bibinfo {author} {\bibfnamefont {M.~P.}\ \bibnamefont
  {Seah}}\ and\ \bibinfo {author} {\bibfnamefont {W.~A.}\ \bibnamefont
  {Dench}},\ }\href {https://doi.org/https://doi.org/10.1002/sia.740010103}
  {\bibfield  {journal} {\bibinfo  {journal} {Surface and Interface Analysis}\
  }\textbf {\bibinfo {volume} {1}},\ \bibinfo {pages} {2} (\bibinfo {year}
  {1979})}\BibitemShut {NoStop}%
\bibitem [{\citenamefont {Takane}\ \emph {et~al.}(2019)\citenamefont {Takane},
  \citenamefont {Wang}, \citenamefont {Souma}, \citenamefont {Nakayama},
  \citenamefont {Nakamura}, \citenamefont {Oinuma}, \citenamefont {Nakata},
  \citenamefont {Iwasawa}, \citenamefont {Cacho}, \citenamefont {Kim},
  \citenamefont {Horiba}, \citenamefont {Kumigashira}, \citenamefont
  {Takahashi}, \citenamefont {Ando},\ and\ \citenamefont
  {Sato}}]{PhysRevLett.122.076402}%
  \BibitemOpen
  \bibfield  {author} {\bibinfo {author} {\bibfnamefont {D.}~\bibnamefont
  {Takane}}, \bibinfo {author} {\bibfnamefont {Z.}~\bibnamefont {Wang}},
  \bibinfo {author} {\bibfnamefont {S.}~\bibnamefont {Souma}}, \bibinfo
  {author} {\bibfnamefont {K.}~\bibnamefont {Nakayama}}, \bibinfo {author}
  {\bibfnamefont {T.}~\bibnamefont {Nakamura}}, \bibinfo {author}
  {\bibfnamefont {H.}~\bibnamefont {Oinuma}}, \bibinfo {author} {\bibfnamefont
  {Y.}~\bibnamefont {Nakata}}, \bibinfo {author} {\bibfnamefont
  {H.}~\bibnamefont {Iwasawa}}, \bibinfo {author} {\bibfnamefont
  {C.}~\bibnamefont {Cacho}}, \bibinfo {author} {\bibfnamefont
  {T.}~\bibnamefont {Kim}}, \bibinfo {author} {\bibfnamefont {K.}~\bibnamefont
  {Horiba}}, \bibinfo {author} {\bibfnamefont {H.}~\bibnamefont {Kumigashira}},
  \bibinfo {author} {\bibfnamefont {T.}~\bibnamefont {Takahashi}}, \bibinfo
  {author} {\bibfnamefont {Y.}~\bibnamefont {Ando}},\ and\ \bibinfo {author}
  {\bibfnamefont {T.}~\bibnamefont {Sato}},\ }\href
  {https://doi.org/10.1103/PhysRevLett.122.076402} {\bibfield  {journal}
  {\bibinfo  {journal} {Phys. Rev. Lett.}\ }\textbf {\bibinfo {volume} {122}},\
  \bibinfo {pages} {076402} (\bibinfo {year} {2019})}\BibitemShut {NoStop}%
\bibitem [{\citenamefont {Rao}\ \emph {et~al.}(2019)\citenamefont {Rao},
  \citenamefont {Li}, \citenamefont {Zhang}, \citenamefont {Tian},
  \citenamefont {Li}, \citenamefont {Fu}, \citenamefont {Tang}, \citenamefont
  {Wang}, \citenamefont {Li}, \citenamefont {Fan}, \citenamefont {Li},
  \citenamefont {Huang}, \citenamefont {Liu}, \citenamefont {Long},
  \citenamefont {Fang}, \citenamefont {Weng}, \citenamefont {Shi},
  \citenamefont {Lei}, \citenamefont {Sun}, \citenamefont {Qian},\ and\
  \citenamefont {Ding}}]{Rao2019}%
  \BibitemOpen
  \bibfield  {author} {\bibinfo {author} {\bibfnamefont {Z.}~\bibnamefont
  {Rao}}, \bibinfo {author} {\bibfnamefont {H.}~\bibnamefont {Li}}, \bibinfo
  {author} {\bibfnamefont {T.}~\bibnamefont {Zhang}}, \bibinfo {author}
  {\bibfnamefont {S.}~\bibnamefont {Tian}}, \bibinfo {author} {\bibfnamefont
  {C.}~\bibnamefont {Li}}, \bibinfo {author} {\bibfnamefont {B.}~\bibnamefont
  {Fu}}, \bibinfo {author} {\bibfnamefont {C.}~\bibnamefont {Tang}}, \bibinfo
  {author} {\bibfnamefont {L.}~\bibnamefont {Wang}}, \bibinfo {author}
  {\bibfnamefont {Z.}~\bibnamefont {Li}}, \bibinfo {author} {\bibfnamefont
  {W.}~\bibnamefont {Fan}}, \bibinfo {author} {\bibfnamefont {J.}~\bibnamefont
  {Li}}, \bibinfo {author} {\bibfnamefont {Y.}~\bibnamefont {Huang}}, \bibinfo
  {author} {\bibfnamefont {Z.}~\bibnamefont {Liu}}, \bibinfo {author}
  {\bibfnamefont {Y.}~\bibnamefont {Long}}, \bibinfo {author} {\bibfnamefont
  {C.}~\bibnamefont {Fang}}, \bibinfo {author} {\bibfnamefont {H.}~\bibnamefont
  {Weng}}, \bibinfo {author} {\bibfnamefont {Y.}~\bibnamefont {Shi}}, \bibinfo
  {author} {\bibfnamefont {H.}~\bibnamefont {Lei}}, \bibinfo {author}
  {\bibfnamefont {Y.}~\bibnamefont {Sun}}, \bibinfo {author} {\bibfnamefont
  {T.}~\bibnamefont {Qian}},\ and\ \bibinfo {author} {\bibfnamefont
  {H.}~\bibnamefont {Ding}},\ }\href
  {https://doi.org/10.1038/s41586-019-1031-8} {\bibfield  {journal} {\bibinfo
  {journal} {Nature}\ }\textbf {\bibinfo {volume} {567}},\ \bibinfo {pages}
  {496} (\bibinfo {year} {2019})}\BibitemShut {NoStop}%
\bibitem [{\citenamefont {Tanaka}\ \emph {et~al.}(2022)\citenamefont {Tanaka},
  \citenamefont {Okazaki}, \citenamefont {Kuroda}, \citenamefont {Noguchi},
  \citenamefont {Arai}, \citenamefont {Minami}, \citenamefont {Ideta},
  \citenamefont {Tanaka}, \citenamefont {Lu}, \citenamefont {Hashimoto},
  \citenamefont {Kandyba}, \citenamefont {Cattelan}, \citenamefont {Barinov},
  \citenamefont {Muro}, \citenamefont {Sasagawa},\ and\ \citenamefont
  {Kondo}}]{PhysRevB.105.L121102}%
  \BibitemOpen
  \bibfield  {author} {\bibinfo {author} {\bibfnamefont {H.}~\bibnamefont
  {Tanaka}}, \bibinfo {author} {\bibfnamefont {S.}~\bibnamefont {Okazaki}},
  \bibinfo {author} {\bibfnamefont {K.}~\bibnamefont {Kuroda}}, \bibinfo
  {author} {\bibfnamefont {R.}~\bibnamefont {Noguchi}}, \bibinfo {author}
  {\bibfnamefont {Y.}~\bibnamefont {Arai}}, \bibinfo {author} {\bibfnamefont
  {S.}~\bibnamefont {Minami}}, \bibinfo {author} {\bibfnamefont
  {S.}~\bibnamefont {Ideta}}, \bibinfo {author} {\bibfnamefont
  {K.}~\bibnamefont {Tanaka}}, \bibinfo {author} {\bibfnamefont
  {D.}~\bibnamefont {Lu}}, \bibinfo {author} {\bibfnamefont {M.}~\bibnamefont
  {Hashimoto}}, \bibinfo {author} {\bibfnamefont {V.}~\bibnamefont {Kandyba}},
  \bibinfo {author} {\bibfnamefont {M.}~\bibnamefont {Cattelan}}, \bibinfo
  {author} {\bibfnamefont {A.}~\bibnamefont {Barinov}}, \bibinfo {author}
  {\bibfnamefont {T.}~\bibnamefont {Muro}}, \bibinfo {author} {\bibfnamefont
  {T.}~\bibnamefont {Sasagawa}},\ and\ \bibinfo {author} {\bibfnamefont
  {T.}~\bibnamefont {Kondo}},\ }\href
  {https://doi.org/10.1103/PhysRevB.105.L121102} {\bibfield  {journal}
  {\bibinfo  {journal} {Phys. Rev. B}\ }\textbf {\bibinfo {volume} {105}},\
  \bibinfo {pages} {L121102} (\bibinfo {year} {2022})}\BibitemShut {NoStop}%
\bibitem [{\citenamefont {Sobota}\ \emph {et~al.}(2021)\citenamefont {Sobota},
  \citenamefont {He},\ and\ \citenamefont {Shen}}]{RevModPhys.93.025006}%
  \BibitemOpen
  \bibfield  {author} {\bibinfo {author} {\bibfnamefont {J.~A.}\ \bibnamefont
  {Sobota}}, \bibinfo {author} {\bibfnamefont {Y.}~\bibnamefont {He}},\ and\
  \bibinfo {author} {\bibfnamefont {Z.-X.}\ \bibnamefont {Shen}},\ }\href
  {https://doi.org/10.1103/RevModPhys.93.025006} {\bibfield  {journal}
  {\bibinfo  {journal} {Rev. Mod. Phys.}\ }\textbf {\bibinfo {volume} {93}},\
  \bibinfo {pages} {025006} (\bibinfo {year} {2021})}\BibitemShut {NoStop}%
\bibitem [{Sup()}]{Supplemental}%
  \BibitemOpen
  \href@noop {} {}\bibinfo {note} {{See Supplemental Material at [URL] for the
  sample preparation and calculation methods and additional data supporting our
  arguments.}}\BibitemShut {Stop}%
\bibitem [{\citenamefont {Muro}\ \emph {et~al.}(2021)\citenamefont {Muro},
  \citenamefont {Senba}, \citenamefont {Ohashi}, \citenamefont {Ohkochi},
  \citenamefont {Matsushita}, \citenamefont {Kinoshita},\ and\ \citenamefont
  {Shin}}]{Muro:ok5049}%
  \BibitemOpen
  \bibfield  {author} {\bibinfo {author} {\bibfnamefont {T.}~\bibnamefont
  {Muro}}, \bibinfo {author} {\bibfnamefont {Y.}~\bibnamefont {Senba}},
  \bibinfo {author} {\bibfnamefont {H.}~\bibnamefont {Ohashi}}, \bibinfo
  {author} {\bibfnamefont {T.}~\bibnamefont {Ohkochi}}, \bibinfo {author}
  {\bibfnamefont {T.}~\bibnamefont {Matsushita}}, \bibinfo {author}
  {\bibfnamefont {T.}~\bibnamefont {Kinoshita}},\ and\ \bibinfo {author}
  {\bibfnamefont {S.}~\bibnamefont {Shin}},\ }\href
  {https://doi.org/10.1107/S1600577521007487} {\bibfield  {journal} {\bibinfo
  {journal} {Journal of Synchrotron Radiation}\ }\textbf {\bibinfo {volume}
  {28}},\ \bibinfo {pages} {1631} (\bibinfo {year} {2021})}\BibitemShut
  {NoStop}%
\bibitem [{\citenamefont {Cucchi}\ \emph {et~al.}(2019)\citenamefont {Cucchi},
  \citenamefont {Guti{\'e}rrez-Lezama}, \citenamefont {Cappelli}, \citenamefont
  {McKeown~Walker}, \citenamefont {Bruno}, \citenamefont {Tenasini},
  \citenamefont {Wang}, \citenamefont {Ubrig}, \citenamefont {Barreteau},
  \citenamefont {Giannini}, \citenamefont {Gibertini}, \citenamefont {Tamai},
  \citenamefont {Morpurgo},\ and\ \citenamefont
  {Baumberger}}]{doi:10.1021/acs.nanolett.8b04534}%
  \BibitemOpen
  \bibfield  {author} {\bibinfo {author} {\bibfnamefont {I.}~\bibnamefont
  {Cucchi}}, \bibinfo {author} {\bibfnamefont {I.}~\bibnamefont
  {Guti{\'e}rrez-Lezama}}, \bibinfo {author} {\bibfnamefont {E.}~\bibnamefont
  {Cappelli}}, \bibinfo {author} {\bibfnamefont {S.}~\bibnamefont
  {McKeown~Walker}}, \bibinfo {author} {\bibfnamefont {F.~Y.}\ \bibnamefont
  {Bruno}}, \bibinfo {author} {\bibfnamefont {G.}~\bibnamefont {Tenasini}},
  \bibinfo {author} {\bibfnamefont {L.}~\bibnamefont {Wang}}, \bibinfo {author}
  {\bibfnamefont {N.}~\bibnamefont {Ubrig}}, \bibinfo {author} {\bibfnamefont
  {C.}~\bibnamefont {Barreteau}}, \bibinfo {author} {\bibfnamefont
  {E.}~\bibnamefont {Giannini}}, \bibinfo {author} {\bibfnamefont
  {M.}~\bibnamefont {Gibertini}}, \bibinfo {author} {\bibfnamefont
  {A.}~\bibnamefont {Tamai}}, \bibinfo {author} {\bibfnamefont {A.~F.}\
  \bibnamefont {Morpurgo}},\ and\ \bibinfo {author} {\bibfnamefont
  {F.}~\bibnamefont {Baumberger}},\ }\href
  {https://doi.org/10.1021/acs.nanolett.8b04534} {\bibfield  {journal}
  {\bibinfo  {journal} {Nano Letters}\ }\textbf {\bibinfo {volume} {19}},\
  \bibinfo {pages} {554} (\bibinfo {year} {2019})}\BibitemShut {NoStop}%
\bibitem [{\citenamefont {Masubuchi}\ \emph {et~al.}(2022)\citenamefont
  {Masubuchi}, \citenamefont {Sakano}, \citenamefont {Tanaka}, \citenamefont
  {Wakafuji}, \citenamefont {Yamamoto}, \citenamefont {Okazaki}, \citenamefont
  {Watanabe}, \citenamefont {Taniguchi}, \citenamefont {Li}, \citenamefont
  {Ejima}, \citenamefont {Sasagawa}, \citenamefont {Ishizaka},\ and\
  \citenamefont {Machida}}]{Masubuchi2022}%
  \BibitemOpen
  \bibfield  {author} {\bibinfo {author} {\bibfnamefont {S.}~\bibnamefont
  {Masubuchi}}, \bibinfo {author} {\bibfnamefont {M.}~\bibnamefont {Sakano}},
  \bibinfo {author} {\bibfnamefont {Y.}~\bibnamefont {Tanaka}}, \bibinfo
  {author} {\bibfnamefont {Y.}~\bibnamefont {Wakafuji}}, \bibinfo {author}
  {\bibfnamefont {T.}~\bibnamefont {Yamamoto}}, \bibinfo {author}
  {\bibfnamefont {S.}~\bibnamefont {Okazaki}}, \bibinfo {author} {\bibfnamefont
  {K.}~\bibnamefont {Watanabe}}, \bibinfo {author} {\bibfnamefont
  {T.}~\bibnamefont {Taniguchi}}, \bibinfo {author} {\bibfnamefont
  {J.}~\bibnamefont {Li}}, \bibinfo {author} {\bibfnamefont {H.}~\bibnamefont
  {Ejima}}, \bibinfo {author} {\bibfnamefont {T.}~\bibnamefont {Sasagawa}},
  \bibinfo {author} {\bibfnamefont {K.}~\bibnamefont {Ishizaka}},\ and\
  \bibinfo {author} {\bibfnamefont {T.}~\bibnamefont {Machida}},\ }\href
  {https://doi.org/10.1038/s41598-022-14845-z} {\bibfield  {journal} {\bibinfo
  {journal} {Scientific Reports}\ }\textbf {\bibinfo {volume} {12}},\ \bibinfo
  {pages} {10936} (\bibinfo {year} {2022})}\BibitemShut {NoStop}%
\bibitem [{\citenamefont {Ozaki}(2003)}]{PhysRevB.67.155108}%
  \BibitemOpen
  \bibfield  {author} {\bibinfo {author} {\bibfnamefont {T.}~\bibnamefont
  {Ozaki}},\ }\href {https://doi.org/10.1103/PhysRevB.67.155108} {\bibfield
  {journal} {\bibinfo  {journal} {Phys. Rev. B}\ }\textbf {\bibinfo {volume}
  {67}},\ \bibinfo {pages} {155108} (\bibinfo {year} {2003})}\BibitemShut
  {NoStop}%
\bibitem [{\citenamefont {Tanaka}\ \emph {et~al.}(2023)\citenamefont {Tanaka},
  \citenamefont {Kuroda},\ and\ \citenamefont {Matsushita}}]{TANAKA2023147297}%
  \BibitemOpen
  \bibfield  {author} {\bibinfo {author} {\bibfnamefont {H.}~\bibnamefont
  {Tanaka}}, \bibinfo {author} {\bibfnamefont {K.}~\bibnamefont {Kuroda}},\
  and\ \bibinfo {author} {\bibfnamefont {T.}~\bibnamefont {Matsushita}},\
  }\href {https://doi.org/https://doi.org/10.1016/j.elspec.2023.147297}
  {\bibfield  {journal} {\bibinfo  {journal} {Journal of Electron Spectroscopy
  and Related Phenomena}\ }\textbf {\bibinfo {volume} {264}},\ \bibinfo {pages}
  {147297} (\bibinfo {year} {2023})}\BibitemShut {NoStop}%
\bibitem [{\citenamefont {Henck}\ \emph {et~al.}(2017)\citenamefont {Henck},
  \citenamefont {Pierucci}, \citenamefont {Fugallo}, \citenamefont {Avila},
  \citenamefont {Cassabois}, \citenamefont {Dappe}, \citenamefont {Silly},
  \citenamefont {Chen}, \citenamefont {Gil}, \citenamefont {Gatti},
  \citenamefont {Sottile}, \citenamefont {Sirotti}, \citenamefont {Asensio},\
  and\ \citenamefont {Ouerghi}}]{PhysRevB.95.085410}%
  \BibitemOpen
  \bibfield  {author} {\bibinfo {author} {\bibfnamefont {H.}~\bibnamefont
  {Henck}}, \bibinfo {author} {\bibfnamefont {D.}~\bibnamefont {Pierucci}},
  \bibinfo {author} {\bibfnamefont {G.}~\bibnamefont {Fugallo}}, \bibinfo
  {author} {\bibfnamefont {J.}~\bibnamefont {Avila}}, \bibinfo {author}
  {\bibfnamefont {G.}~\bibnamefont {Cassabois}}, \bibinfo {author}
  {\bibfnamefont {Y.~J.}\ \bibnamefont {Dappe}}, \bibinfo {author}
  {\bibfnamefont {M.~G.}\ \bibnamefont {Silly}}, \bibinfo {author}
  {\bibfnamefont {C.}~\bibnamefont {Chen}}, \bibinfo {author} {\bibfnamefont
  {B.}~\bibnamefont {Gil}}, \bibinfo {author} {\bibfnamefont {M.}~\bibnamefont
  {Gatti}}, \bibinfo {author} {\bibfnamefont {F.}~\bibnamefont {Sottile}},
  \bibinfo {author} {\bibfnamefont {F.}~\bibnamefont {Sirotti}}, \bibinfo
  {author} {\bibfnamefont {M.~C.}\ \bibnamefont {Asensio}},\ and\ \bibinfo
  {author} {\bibfnamefont {A.}~\bibnamefont {Ouerghi}},\ }\href
  {https://doi.org/10.1103/PhysRevB.95.085410} {\bibfield  {journal} {\bibinfo
  {journal} {Phys. Rev. B}\ }\textbf {\bibinfo {volume} {95}},\ \bibinfo
  {pages} {085410} (\bibinfo {year} {2017})}\BibitemShut {NoStop}%
\bibitem [{\citenamefont {Di~Sante}\ \emph {et~al.}(2017)\citenamefont
  {Di~Sante}, \citenamefont {Das}, \citenamefont {Bigi}, \citenamefont
  {Erg\"onenc}, \citenamefont {G\"urtler}, \citenamefont {Krieger},
  \citenamefont {Schmitt}, \citenamefont {Ali}, \citenamefont {Rossi},
  \citenamefont {Thomale}, \citenamefont {Franchini}, \citenamefont {Picozzi},
  \citenamefont {Fujii}, \citenamefont {Strocov}, \citenamefont {Sangiovanni},
  \citenamefont {Vobornik}, \citenamefont {Cava},\ and\ \citenamefont
  {Panaccione}}]{PhysRevLett.119.026403}%
  \BibitemOpen
  \bibfield  {author} {\bibinfo {author} {\bibfnamefont {D.}~\bibnamefont
  {Di~Sante}}, \bibinfo {author} {\bibfnamefont {P.~K.}\ \bibnamefont {Das}},
  \bibinfo {author} {\bibfnamefont {C.}~\bibnamefont {Bigi}}, \bibinfo {author}
  {\bibfnamefont {Z.}~\bibnamefont {Erg\"onenc}}, \bibinfo {author}
  {\bibfnamefont {N.}~\bibnamefont {G\"urtler}}, \bibinfo {author}
  {\bibfnamefont {J.~A.}\ \bibnamefont {Krieger}}, \bibinfo {author}
  {\bibfnamefont {T.}~\bibnamefont {Schmitt}}, \bibinfo {author} {\bibfnamefont
  {M.~N.}\ \bibnamefont {Ali}}, \bibinfo {author} {\bibfnamefont
  {G.}~\bibnamefont {Rossi}}, \bibinfo {author} {\bibfnamefont
  {R.}~\bibnamefont {Thomale}}, \bibinfo {author} {\bibfnamefont
  {C.}~\bibnamefont {Franchini}}, \bibinfo {author} {\bibfnamefont
  {S.}~\bibnamefont {Picozzi}}, \bibinfo {author} {\bibfnamefont
  {J.}~\bibnamefont {Fujii}}, \bibinfo {author} {\bibfnamefont {V.~N.}\
  \bibnamefont {Strocov}}, \bibinfo {author} {\bibfnamefont {G.}~\bibnamefont
  {Sangiovanni}}, \bibinfo {author} {\bibfnamefont {I.}~\bibnamefont
  {Vobornik}}, \bibinfo {author} {\bibfnamefont {R.~J.}\ \bibnamefont {Cava}},\
  and\ \bibinfo {author} {\bibfnamefont {G.}~\bibnamefont {Panaccione}},\
  }\href {https://doi.org/10.1103/PhysRevLett.119.026403} {\bibfield  {journal}
  {\bibinfo  {journal} {Phys. Rev. Lett.}\ }\textbf {\bibinfo {volume} {119}},\
  \bibinfo {pages} {026403} (\bibinfo {year} {2017})}\BibitemShut {NoStop}%
\bibitem [{\citenamefont {Matsui}\ \emph {et~al.}(2018)\citenamefont {Matsui},
  \citenamefont {Nishikawa}, \citenamefont {Daimon}, \citenamefont {Muntwiler},
  \citenamefont {Takizawa}, \citenamefont {Namba},\ and\ \citenamefont
  {Greber}}]{PhysRevB.97.045430}%
  \BibitemOpen
  \bibfield  {author} {\bibinfo {author} {\bibfnamefont {F.}~\bibnamefont
  {Matsui}}, \bibinfo {author} {\bibfnamefont {H.}~\bibnamefont {Nishikawa}},
  \bibinfo {author} {\bibfnamefont {H.}~\bibnamefont {Daimon}}, \bibinfo
  {author} {\bibfnamefont {M.}~\bibnamefont {Muntwiler}}, \bibinfo {author}
  {\bibfnamefont {M.}~\bibnamefont {Takizawa}}, \bibinfo {author}
  {\bibfnamefont {H.}~\bibnamefont {Namba}},\ and\ \bibinfo {author}
  {\bibfnamefont {T.}~\bibnamefont {Greber}},\ }\href
  {https://doi.org/10.1103/PhysRevB.97.045430} {\bibfield  {journal} {\bibinfo
  {journal} {Phys. Rev. B}\ }\textbf {\bibinfo {volume} {97}},\ \bibinfo
  {pages} {045430} (\bibinfo {year} {2018})}\BibitemShut {NoStop}%
\bibitem [{\citenamefont {Strocov}(2003)}]{STROCOV200365}%
  \BibitemOpen
  \bibfield  {author} {\bibinfo {author} {\bibfnamefont {V.}~\bibnamefont
  {Strocov}},\ }\href
  {https://doi.org/https://doi.org/10.1016/S0368-2048(03)00054-9} {\bibfield
  {journal} {\bibinfo  {journal} {Journal of Electron Spectroscopy and Related
  Phenomena}\ }\textbf {\bibinfo {volume} {130}},\ \bibinfo {pages} {65}
  (\bibinfo {year} {2003})}\BibitemShut {NoStop}%
\bibitem [{\citenamefont {Moser}(2017)}]{MOSER201729}%
  \BibitemOpen
  \bibfield  {author} {\bibinfo {author} {\bibfnamefont {S.}~\bibnamefont
  {Moser}},\ }\href
  {https://doi.org/https://doi.org/10.1016/j.elspec.2016.11.007} {\bibfield
  {journal} {\bibinfo  {journal} {Journal of Electron Spectroscopy and Related
  Phenomena}\ }\textbf {\bibinfo {volume} {214}},\ \bibinfo {pages} {29}
  (\bibinfo {year} {2017})}\BibitemShut {NoStop}%
\bibitem [{\citenamefont {Kanter}(1970)}]{PhysRevB.1.522}%
  \BibitemOpen
  \bibfield  {author} {\bibinfo {author} {\bibfnamefont {H.}~\bibnamefont
  {Kanter}},\ }\href {https://doi.org/10.1103/PhysRevB.1.522} {\bibfield
  {journal} {\bibinfo  {journal} {Phys. Rev. B}\ }\textbf {\bibinfo {volume}
  {1}},\ \bibinfo {pages} {522} (\bibinfo {year} {1970})}\BibitemShut {NoStop}%
\bibitem [{\citenamefont {Klasson}\ \emph {et~al.}(1972)\citenamefont
  {Klasson}, \citenamefont {Hedman}, \citenamefont {Berndtsson}, \citenamefont
  {Nilsson}, \citenamefont {Nordling},\ and\ \citenamefont
  {Melnik}}]{M_Klasson_1972}%
  \BibitemOpen
  \bibfield  {author} {\bibinfo {author} {\bibfnamefont {M.}~\bibnamefont
  {Klasson}}, \bibinfo {author} {\bibfnamefont {J.}~\bibnamefont {Hedman}},
  \bibinfo {author} {\bibfnamefont {A.}~\bibnamefont {Berndtsson}}, \bibinfo
  {author} {\bibfnamefont {R.}~\bibnamefont {Nilsson}}, \bibinfo {author}
  {\bibfnamefont {C.}~\bibnamefont {Nordling}},\ and\ \bibinfo {author}
  {\bibfnamefont {P.}~\bibnamefont {Melnik}},\ }\href
  {https://doi.org/10.1088/0031-8949/5/1-2/015} {\bibfield  {journal} {\bibinfo
   {journal} {Physica Scripta}\ }\textbf {\bibinfo {volume} {5}},\ \bibinfo
  {pages} {93} (\bibinfo {year} {1972})}\BibitemShut {NoStop}%
\bibitem [{\citenamefont {Liu}\ \emph {et~al.}(2014)\citenamefont {Liu},
  \citenamefont {Zhou}, \citenamefont {Zhang}, \citenamefont {Wang},
  \citenamefont {Weng}, \citenamefont {Prabhakaran}, \citenamefont {Mo},
  \citenamefont {Shen}, \citenamefont {Fang}, \citenamefont {Dai},
  \citenamefont {Hussain},\ and\ \citenamefont
  {Chen}}]{doi:10.1126/science.1245085}%
  \BibitemOpen
  \bibfield  {author} {\bibinfo {author} {\bibfnamefont {Z.~K.}\ \bibnamefont
  {Liu}}, \bibinfo {author} {\bibfnamefont {B.}~\bibnamefont {Zhou}}, \bibinfo
  {author} {\bibfnamefont {Y.}~\bibnamefont {Zhang}}, \bibinfo {author}
  {\bibfnamefont {Z.~J.}\ \bibnamefont {Wang}}, \bibinfo {author}
  {\bibfnamefont {H.~M.}\ \bibnamefont {Weng}}, \bibinfo {author}
  {\bibfnamefont {D.}~\bibnamefont {Prabhakaran}}, \bibinfo {author}
  {\bibfnamefont {S.-K.}\ \bibnamefont {Mo}}, \bibinfo {author} {\bibfnamefont
  {Z.~X.}\ \bibnamefont {Shen}}, \bibinfo {author} {\bibfnamefont
  {Z.}~\bibnamefont {Fang}}, \bibinfo {author} {\bibfnamefont {X.}~\bibnamefont
  {Dai}}, \bibinfo {author} {\bibfnamefont {Z.}~\bibnamefont {Hussain}},\ and\
  \bibinfo {author} {\bibfnamefont {Y.~L.}\ \bibnamefont {Chen}},\ }\href
  {https://doi.org/10.1126/science.1245085} {\bibfield  {journal} {\bibinfo
  {journal} {Science}\ }\textbf {\bibinfo {volume} {343}},\ \bibinfo {pages}
  {864} (\bibinfo {year} {2014})}\BibitemShut {NoStop}%
\end{thebibliography}%


\end{document}

\title{Supplemental Material: Broken Screw Rotational Symmetry \\ in the Near-Surface Electronic Structure of $AB$-Stacked Crystals}

\author{Hiroaki~Tanaka}
\affiliation{Institute for Solid State Physics, The University of Tokyo, Kashiwa, Chiba 277-8581, Japan}

\author{Shota~Okazaki}
\affiliation{Materials and Structures Laboratory, Tokyo Institute of Technology, Yokohama, Kanagawa 226-8503, Japan}

\author{Masaru~Kobayashi}
\affiliation{Materials and Structures Laboratory, Tokyo Institute of Technology, Yokohama, Kanagawa 226-8503, Japan}

\author{Yuto~Fukushima}
\affiliation{Institute for Solid State Physics, The University of Tokyo, Kashiwa, Chiba 277-8581, Japan}

\author{Yosuke~Arai}
\affiliation{Institute for Solid State Physics, The University of Tokyo, Kashiwa, Chiba 277-8581, Japan}

\author{Takushi~Iimori}
\affiliation{Institute for Solid State Physics, The University of Tokyo, Kashiwa, Chiba 277-8581, Japan}

\author{Mikk~Lippmaa}
\affiliation{Institute for Solid State Physics, The University of Tokyo, Kashiwa, Chiba 277-8581, Japan}

\author{Kohei~Yamagami}
\affiliation{Japan Synchrotron Radiation Research Institute (JASRI), Sayo, Hyogo 679-5198, Japan}

\author{Yoshinori~Kotani}
\affiliation{Japan Synchrotron Radiation Research Institute (JASRI), Sayo, Hyogo 679-5198, Japan}

\author{Fumio~Komori}
\affiliation{Institute of Industrial Science, The University of Tokyo, Meguro-ku, Tokyo 153-8505, Japan}

\author{Kenta~Kuroda}
%\email{kuroken224@hiroshima-u.ac.jp}
\affiliation{Graduate School of Advanced Science and Engineering, Hiroshima University, Higashi-hiroshima, Hiroshima 739-8526, Japan}
\affiliation{International Institute for Sustainability with Knotted Chiral Meta Matter (WPI-SKCM${}^{2}$), Higashi-hiroshima, Hiroshima 739-8526, Japan}

\author{Takao~Sasagawa}
%\email{sasagawa@msl.titech.ac.jp}
\affiliation{Materials and Structures Laboratory, Tokyo Institute of Technology, Yokohama, Kanagawa 226-8503, Japan}

\author{Takeshi~Kondo}
%\email{kondo1215@issp.u-tokyo.ac.jp}
\affiliation{Institute for Solid State Physics, The University of Tokyo, Kashiwa, Chiba 277-8581, Japan}
\affiliation{Trans-scale Quantum Science Institute, The University of Tokyo, Bunkyo-ku, Tokyo 113-0033, Japan}

\date{\today}

\maketitle

\renewcommand{\thefigure}{S\arabic{figure}}
\renewcommand{\thetable}{S\Roman{table}}
\titleformat*{\section}{\bfseries}

\section{Note 1: Single crystal growth and thin flake fabrication of $h$-BN}
\begin{figure}[hb]
\centering
\includegraphics{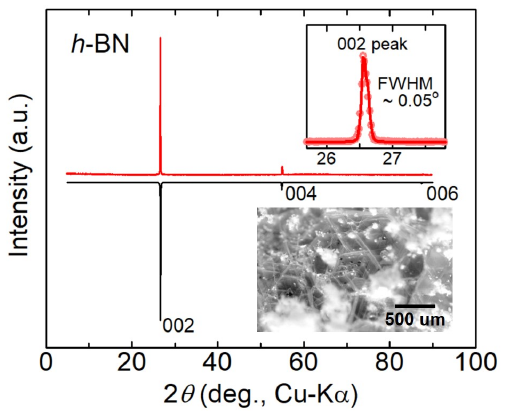}
\caption{\label{Fig: hBN_characterization} X-ray diffraction (XRD) pattern from the cleaved surface of an $h$-BN crystal. The red and black curves represent the experimental and simulation spectra, respectively. Lower inset: the grown single crystals of $h$-BN. Upper inset: the $(002)$ peak of the XRD pattern. }
\end{figure}

Using $h$-BN powder, $h$-BN crucibles, and the Ni-Cr-alloy flux, single crystals of $h$-BN were grown under atmospheric pressure at 1500 -- 1550~$^\circ$C  in a tube furnace with a nitrogen gas flow rate of 0.2 L/min, yielding transparent single crystals of about 0.5 mm in diameter [lower inset of Fig.\ \ref{Fig: hBN_characterization}]. As shown in Fig.\ \ref{Fig: hBN_characterization}, only $(00l)$ peaks were observed in the X-ray diffraction pattern from the cleaved surface of an $h$-BN crystal. The full width at half maximum of the (002) peak was 0.05${}^\circ$ [upper inset of Fig. \ref{Fig: hBN_characterization}], indicating the high crystallinity of the grown crystal. 

We exfoliated $h$-BN single crystals onto a $\mathrm{Si}\mathrm{O}_2/\mathrm{Si}$ substrate to obtain thin flakes.
Large flakes with a width of several tens of micrometers were picked up with a polydimethylsiloxane and polycarbonate polymer stamp and dropped onto a few-layer graphene substrate grown on $4H$-$\mathrm{Si}\mathrm{C}$.
Since the unit cells of $h$-BN and graphene are approximately the same, graphite is a suitable conductive substrate for $h$-BN and vice versa [20].
Graphene layers were fabricated by annealing $4H$-$\mathrm{Si}\mathrm{C}$ substrate at about $1350{}^\circ\mathrm{C}$ for 20 minutes in a vacuum better than $6\times10^{-7}\ \textrm{Pa}$.
We performed these exfoliation and pickup procedures in a nitrogen-filled glovebox to avoid surface damage of $h$-BN flakes.
After that, $h$-BN flakes on graphene/SiC substrates were transported to the ARPES chamber of SPring-8 BL25SU in a vacuum suitcase.
During transport, a non-evaporable getter kept the vacuum in the suitcase at around $6\times10^{-5}$~Pa.
The samples were thus never exposed to ambient air, which can easily damage the sample surface and degrade ARPES spectra.
Before ARPES measurements, the samples were annealed at $400{}^\circ\mathrm{C}$ for two hours in the vacuum chamber to clean the surface.

\begin{figure}
\centering
\includegraphics{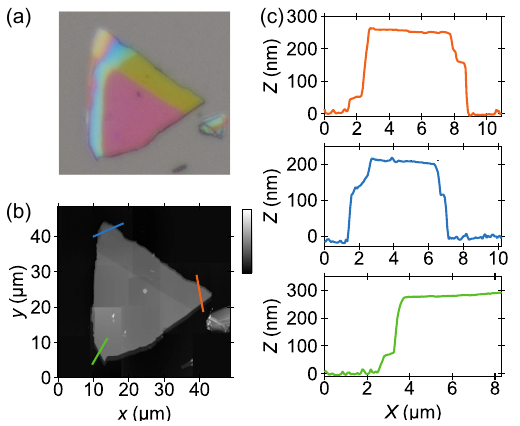}
\caption{\label{Fig: hBN_sample} Microscope images of an $h$-BN flake. (a)~Optical microscope image. The $k_x$ axis in Fig.\ 1(b) corresponds to the vertical direction in this panel. (b)~Atomic force microscope image obtained by stitching 20-micrometer-wide images. (c)~Height profiles along the lines in (b).}
\end{figure}

The sample thickness was later examined with tapping-mode atomic force microscopy in the air.
Figure \ref{Fig: hBN_sample} shows the $h$-BN flake from which we successfully obtained the band dispersion.
The flake thickness was between 200 and 300~nm, which is small enough to prevent surface charge up during the ARPES experiment, but large enough to avoid electronic structure quantization.

\clearpage

\section{Note 2: Band dispersion and photoemission intensity calculations}
We calculated the photoemission intensity by SPADExp [19], which uses the ground state wave functions obtained from OpenMX [18].
We employed a thick slab system with dozens of bilayers (BLs) to simulate the near-surface electronic structure of a bulk crystal.
For $2H$-$\mathrm{Nb}\mathrm{S}_2$, the slab included 20 BLs when spin-orbit coupling (SOC) was neglected and 10 BLs when SOC was included.
For $h$-BN, we used a 40 BL slab, and SOC was neglected.

\begin{figure}[hb]
\centering
\includegraphics{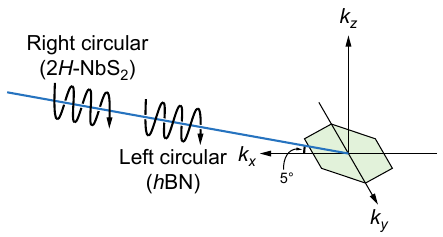}
\caption{\label{Fig: ARPES_orientation} Light polarization and incident angle in SX-ARPES measurements.}
\end{figure}

We used the same incident light electric field direction condition as in ARPES experiments [Fig.\ \ref{Fig: ARPES_orientation}].
Since both crystals have two possible termination surfaces, we show the photoemission intensity distribution summed up for these two termination conditions.

The energy axes of the calculated results were shifted by $-0.3\ \mathrm{eV}$ and $-1\ \mathrm{eV}$ for $2H$-$\mathrm{Nb}\mathrm{S}_2$ and $h$-BN, respectively, to adjust the band position to experimental results.
Such an energy shift does not affect our discussion because we only focus on symmetry-driven band degeneracy and its disappearance near the surface.
We note that the overall band shape of calculations coincides well with experiments.

\section{Note 3: Supporting data of SX-ARPES measurements of $2H$-$\mathbf{Nb}\mathbf{S}_2$}
\begin{figure}[hb]
\centering
\includegraphics{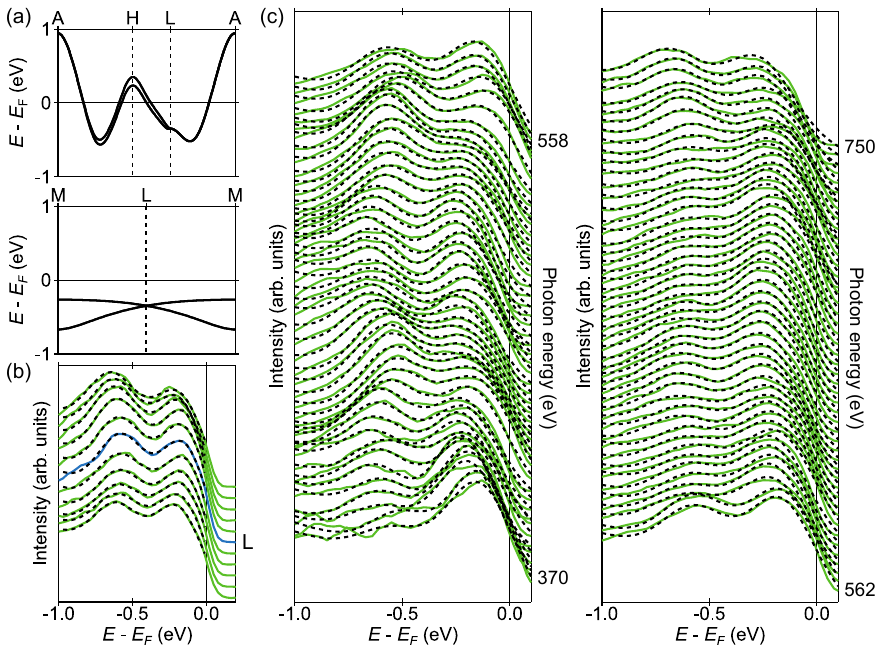}
\caption{\label{Fig: NbS2_EDCs} Band dispersion and energy distribution curves (EDCs) of $2H$-$\mathrm{Nb}\mathrm{S}_2$. (a)~Bulk band dispersion along the high symmetry path on the $k_z=\pi/c$ plane and the $ML$ path. Spin-orbit coupling was included in the calculations. (b)~EDCs around the $L$ point taken with 525 eV SX light, corresponding to Fig.\ 1(c). (c)~Photon energy dependence of the EDC at the $\bar{M}$ point, corresponding to Fig.\ 1(e).  In (b) and (c) the black dashed curves represent the fitting results.}
\end{figure}

In the bulk band dispersion of $2H$-$\mathrm{Nb}\mathrm{S}_2$, a nodal plane on $k_z=\pi/c$ can exhibit an energy gap due to the inclusion of spin-orbit coupling.
However, band degeneracy partially remains, for example along the $AL$ path [Fig.\ \ref{Fig: NbS2_EDCs}(a)].

We used the fitting function comprising one or two Gaussians and a linear background in all curve fitting analyses represented below.
For the conversion of the photon energy to the momentum, we used the work function and inner potential parameters $W=5\ \mathrm{eV}$ and $V_0=10\ \mathrm{eV}$ respectively.
They are the same as those in the previous study on $2H$-$\mathrm{Nb}\mathrm{S}_2$ [12], showing a good symmetric property with respect to the $M$ point for the valence bands around $E_F-2\ \mathrm{eV}$.

\clearpage

\section{Note 4: Core-level spectra}
\begin{figure}[hb]
\centering
\includegraphics{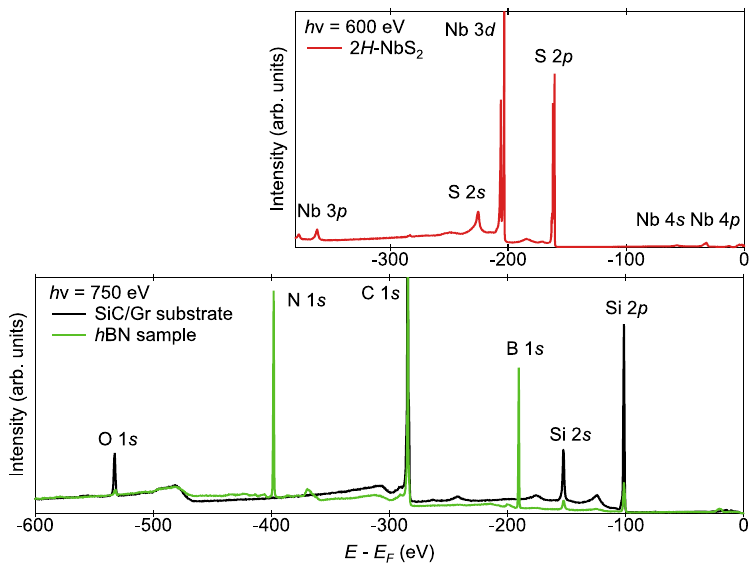}
\caption{\label{Fig: hBN_XPS} X-ray photoemission spectra of $2H$-$\mathrm{Nb}\mathrm{S}_2$, $h$-BN, and SiC/Graphene substrate.}
\end{figure}

Figure \ref{Fig: hBN_XPS} shows the core-level spectra of $2H$-$\mathrm{Nb}\mathrm{S}_2$, $h$-BN, and SiC/Graphene substrate.
Comparing the latter two spectra, we found boron and nitrogen $1s$ peaks only when the soft-x-ray beam was on the $h$-BN flake.
The carbon $1s$ peak also appeared in the $h$-BN spectra, which may be due to the polymer stamp remaining on the flake surface.
We note that the remaining carbon atoms on the $h$-BN surface did not affect our ARPES spectra.

\section{Note 5: Crossover from bulk spectra to surface spectra in SX-ARPES of $h$-BN}

\begin{figure}[hb]
\centering
\includegraphics{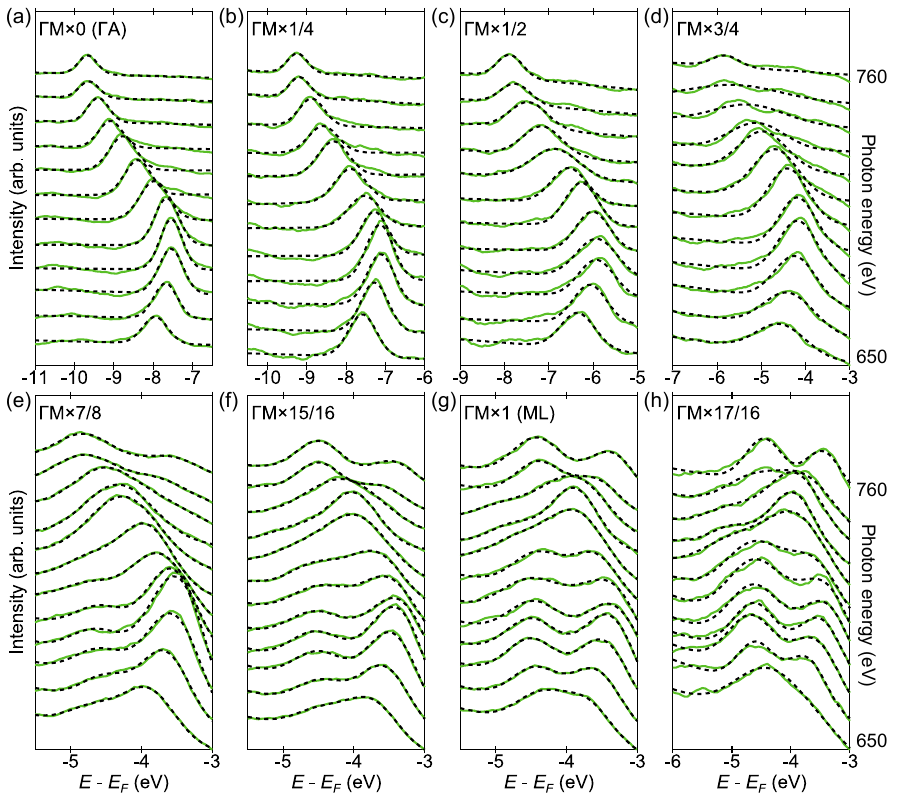}
\caption{\label{Fig: hBN_EDCs_GtoM} Energy distribution curves at several $k_x$ positions and photon energies varying from 650 to 760 eV. The left top equation in each panel represents the distance to the $\Gamma A$ axis, where $\Gamma\mathrm{M}=2\pi/\sqrt{3}a$ is the distance between the $\Gamma$ and $M$ points. The dashed black curves represent the fitting results.}
\end{figure}

\begin{figure}[ht]
\centering
\includegraphics{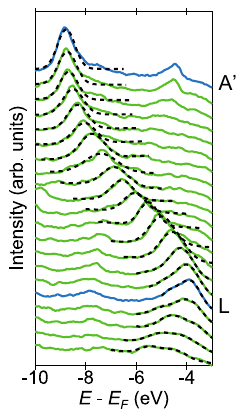}
\caption{\label{Fig: hBN_EDCs_720} Energy distribution curves around the $L$ point taken with 720 eV SX light, corresponding to Fig.\ 2(c). The black dashed curves represent the fitting results.}
\end{figure}

Figure \ref{Fig: hBN_EDCs_GtoM} summarizes the photon energy dependence of energy distribution curves (EDCs) at several $k_x$ positions.
In the small $k_x$ region, we observed one symmetric peak in each EDC, fitted with a sum of a single Gaussian and a linear background.
When the $k_x$ coordinate is closer to the $M$ point, a second peak appears.
Therefore, the spectra were fitted with two Gaussians and a linear background.
The number of Gaussians used in the fitting is also applied to the EDC analysis of the band dispersion taken with 720~eV SX light [Fig.\ \ref{Fig: hBN_EDCs_720}].

\begin{figure}[hb]
\centering
\includegraphics{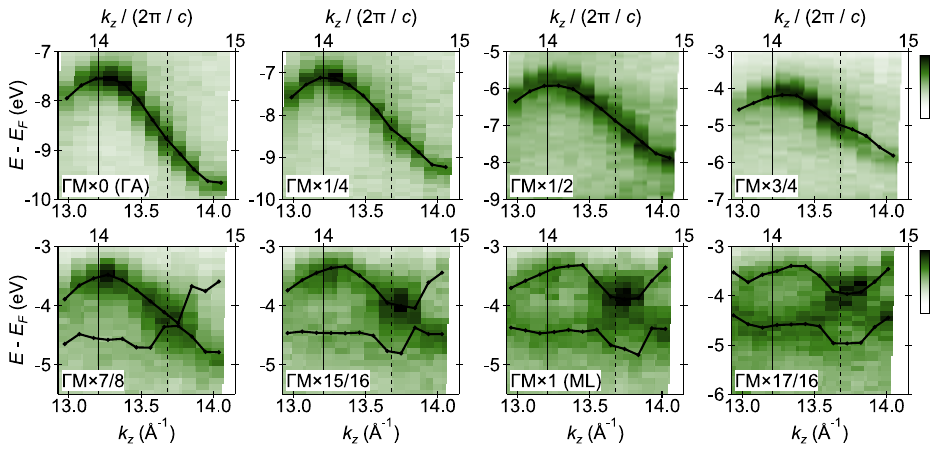}
\caption{\label{Fig: hBN_kz_GtoM} $k_z$ dispersions at several $k_x$ positions. The black curves represent peak positions determined by the fitting analysis, as described above. }
\end{figure}

The fitting results are shown on the $k_z$ dispersion maps in Fig.\ \ref{Fig: hBN_kz_GtoM}.
We used the inner potential and work function values $V_0=6.33\ \mathrm{eV}$ and $W=5\ \mathrm{eV}$ respectively so that the band dispersion along the $\Gamma A$ path becomes symmetric with respect to the $\Gamma$ point ($k_z=14\times 2\pi/c$).
These results indicate that the second peak growth and the crossover to the gapped surface spectra simultaneously happen at about the $\Gamma\mathrm{M}\times7/8$ position.

\clearpage
\section{Note 6: $k_z$-broadening simulations}
We performed the $k_z$-broadening simulations following the previous study [23] and observed the band position shift when the Lorentzian broadening was substantial.
However, the typical broadening length scale $\Delta k_z = 0.1\ \text{\AA}^{-1}$ was insufficient to cause the band splitting. 

In our simulations, we used two cosine bands $E=\pm 0.5\cos(ck/2)$ [eV].
For the energy broadening in the initial states, we employed the Gaussian broadening $\exp(-1/2\cdot ((E-E_0)/\Delta E)^2)$ with $\Delta E=80\ \text{meV}$ reflecting the energy resolution of our ARPES experiments.
We did not use the Lorentzian form because the imaginary self-energy related to the Lorentzian broadening will be negligible for weakly correlated $h$-BN.
For the wavevector broadening in the final states, we employed the Lorentzian broadening $1/((k-k_0 )^2+(\Delta k/2)^2)$, associated with the exponential wave function decay $\exp(-\Delta k\cdot x/2)=\exp(-x / 2\lambda)$; 2 appears in the denominator because the square of the decay, related with the scattering probability, has the form $\exp(-\Delta k\cdot x)=\exp(-x/\lambda)$.

\begin{figure}[hb]
\centering
\includegraphics{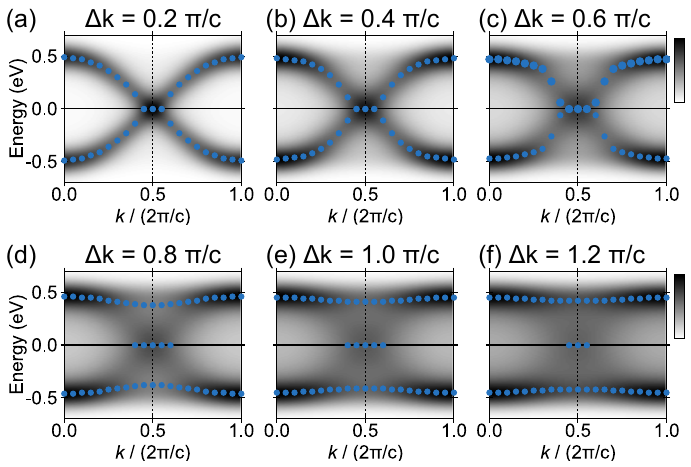}
\caption{\label{Fig: Broadening_simulation} $k_z$-broadening simulations for different $\Delta k$ values. The blue dots represent local maxima extracted from energy distribution curves. }
\end{figure}

Fig.\ \ref{Fig: Broadening_simulation} summarizes the result.
The deviation from the original cosine bands grew with increasing the broadening width $\Delta k$.
The degeneration property was retained up to $\Delta k=0.6 \pi/c$.
Our result is consistent with the previous study [23]; when $\Delta k=0.6 |\Gamma X|$, their simulation shows energy position shift although the overall shape is kept.
When $\Delta k$ is larger than $0.8 \pi/c$, the degeneracy is broken, resulting in weakly dispersing gapped two bands.
Since the observed energy maxima at $E=0$ are very small, they are unimportant for our discussion.

The $k_z$-broadening scheme can also explain the band degeneracy breaking.
However, the Lorentzian broadening $\Delta k = 0.1\ \text{\AA}^{-1}$, typical length scale for soft-x-ray ARPES [13], corresponds to $0.2 \pi/c$ for $h$-BN and $0.4 \pi/c$ for $2H$-$\mathrm{Nb}\mathrm{S}_2$.
Therefore, the $k_z$-broadening simulation cannot explain the gap opening with such a small $\Delta k$ value.
The decay function shape or its length scale needs to be reconsidered to explain the experimental results.

\section{Note 7: Photoemission intensity calculations of $h$-BN bulk}

\begin{figure}[hb]
\centering
\includegraphics{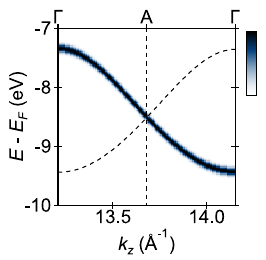}
\caption{\label{Fig: hBN_PAD_bulk_GA} Calculated photoemission intensity of $h$-BN bulk along the $\Gamma A$ path. The color map is normalized so that the integrated intensity along the energy becomes uniform. The dashed curves represent the bulk band dispersion.}
\end{figure}

Figure \ref{Fig: hBN_PAD_bulk_GA} shows the calculated photoemission intensity of $h$-BN bulk.
Although the bulk band dispersion includes a pair of bands, one has sufficient intensity, whereas the other is too weak to be seen in the dispersion plot.
This phenomenon frequently happens in the $k_z$ dispersion of $AB$-stacked crystals, as discussed in the main text.
We note that we got the same result as ARPES measurements [Fig.\ 2(e)] in terms of which band has strong intensity.

\clearpage

\section{Note 8: Crossover from bulk spectra to surface spectra in calculations of $h$-BN}
\begin{figure}[hb]
\centering
\includegraphics{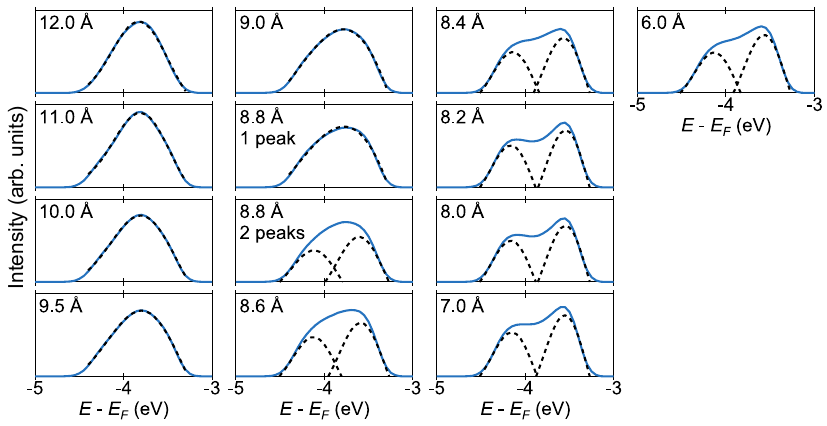}
\caption{\label{Fig: hBN_PAD_slab_depth} Calculated photoemission intensity of $h$-BN at the $L$ point using the linear decay function. The black dashed curves represent the peaks included in the fitting function.}
\end{figure}

Figure \ref{Fig: hBN_PAD_slab_depth} summarizes the photoemission calculation results using the linear decay function.
When the depth parameter $\lambda$ is large, the spectra at the $L$ point exhibit a single peak.
The residual error of the one-peak fitting becomes noticeable at around $\lambda=8.8\ \text{\AA}$.
Therefore, we fitted the spectra with one Gaussian when $\lambda\ge 8.8\ \text{\AA}$ and with two Gaussians when $\lambda\le 8.8\ \text{\AA}$.

\clearpage

\section{Note 9: Energy distribution curves of calculated photoemission angular distributions of $h$-BN}
\begin{figure}[hb]
\centering
\includegraphics{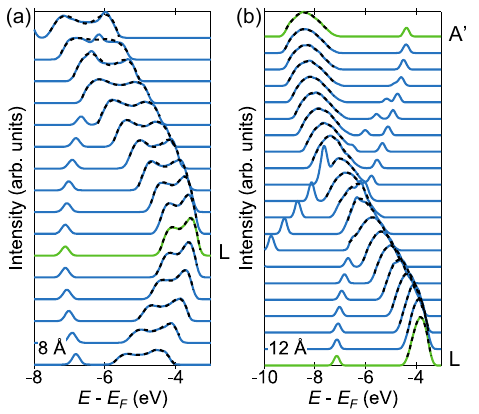}
\caption{\label{Fig: hBN_PAD_EDCs_AL} Energy distribution curves of calculated photoemission intensity along the path crossing the $L$ points, corresponding to Figs.\ 4(a) and 4(b). The dashed curves represent the fitting functions.}
\end{figure}

\begin{figure}[hb]
\centering
\includegraphics{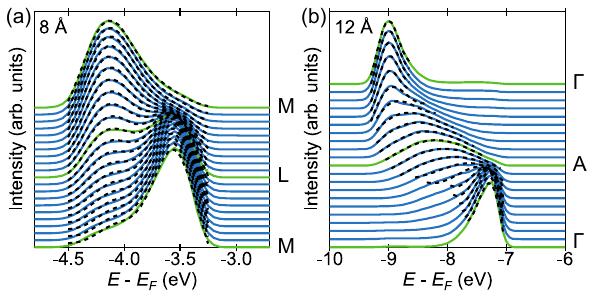}
\caption{\label{Fig: hBN_PAD_EDCs_kz} Energy distribution curves of calculated photoemission intensity along the $ML$ and $\Gamma A$ paths, corresponding to Figs.\ 4(c) and 4(d). The dashed curves represent the fitting functions.}
\end{figure}

When fitting the calculated energy distribution curves [Figs.\ \ref{Fig: hBN_PAD_EDCs_AL} and \ref{Fig: hBN_PAD_EDCs_kz}], a single Gaussian was used for $\lambda=12\ \text{\AA}$ and two Gaussians for $\lambda=8\ \text{\AA}$.

\clearpage

\section{Note 10: Photoemission angular distribution calculations of $2H$-$\mathbf{Nb}\mathbf{S}_2$}
For all cases, two Gaussian components were used for fitting the calculated energy distribution curves [Fig.\ \ref{Fig: NbS2_PAD_EDCs_AL}].

\begin{figure}[hb]
\centering
\includegraphics{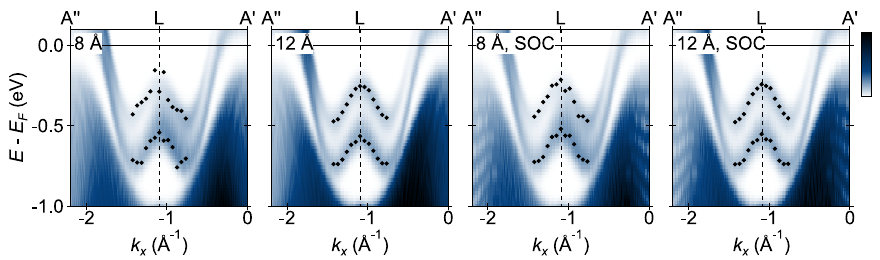}
\caption{\label{Fig: NbS2_PAD_maps} Calculated band dispersions of $2H$-$\mathrm{Nb}\mathrm{S}_2$ along the path crossing the $L$ point. The black dots represent peak positions determined by fitting the energy distribution curves [Fig.\ \ref{Fig: NbS2_PAD_EDCs_AL}]. The left top labels represent the decay length parameter $\lambda$ and whether spin-orbit coupling was included.}
\end{figure}

\begin{figure}[hb]
\centering
\includegraphics{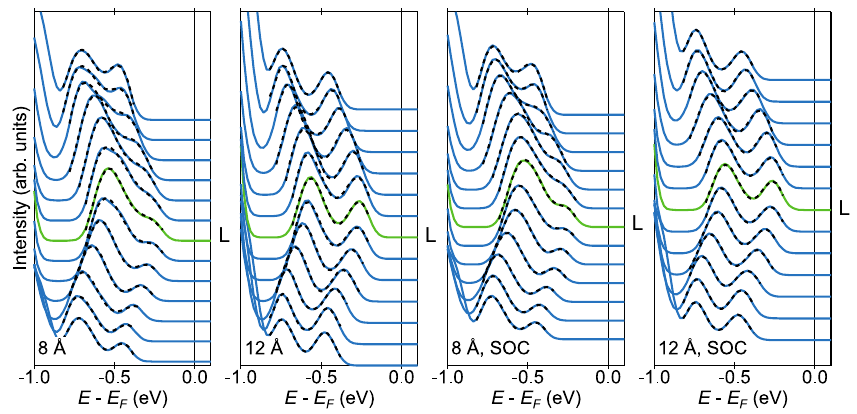}
\caption{\label{Fig: NbS2_PAD_EDCs_AL} Energy distribution curves of calculated photoemission intensity around the $L$ point, corresponding to Fig.\ \ref{Fig: NbS2_PAD_maps}. The dashed curves represent the fitting functions.}
\end{figure}